\documentclass[11pt]{article}
\usepackage{graphicx}
\usepackage{amsfonts}
\usepackage{epstopdf}
\usepackage{float}
\newcommand{\compl}{{\mathbb C}}
\newcommand{\real}{{\mathbb R}}
\newcommand{\captionfonts}{\footnotesize}
\makeatletter  % Allow the use of @ in command names
\long\def\@makecaption#1#2{%
  \vskip\abovecaptionskip
  \sbox\@tempboxa{{\captionfonts #1: #2}}%
  \ifdim \wd\@tempboxa >\hsize
    {\captionfonts #1: #2\par}
  \else
    \hbox to\hsize{\hfil\box\@tempboxa\hfil}%
  \fi
  \vskip\belowcaptionskip}
\makeatother   % Cancel the effect of \makeatletter
\begin{document}
\title{Quantum Interference and Superposition in Cognition: A Theory for the Disjunction of Concepts}
\author{Diederik Aerts\\
        \normalsize\itshape
        Center Leo Apostel for Interdisciplinary Studies \\
        \normalsize\itshape
        and Departments of Mathematics and Psychology \\
        \normalsize\itshape
        Vrije Universiteit Brussel, 1160 Brussels, 
       Belgium \\
        \normalsize
        E-Mail: \textsf{diraerts@vub.ac.be}
        }
\date{}
\maketitle
\begin{abstract}
\noindent
We elaborate a theory for the modeling of concepts using the mathematical structure of quantum mechanics. Concepts are represented by vectors in the complex Hilbert space of quantum mechanics and membership weights of items are modeled by quantum weights calculated following the quantum rules. We apply this theory to model the disjunction of concepts and show that experimental data of membership weights of items with respect to the disjunction of concepts can be modeled accurately. It is the quantum effects of interference and superposition, combined with an effect of context, that are at the origin of the effects of overextension and underextension observed as deviations from a classical use of the disjunction. We put forward a graphical explanation of the effects of overextension and underextension by interpreting the quantum model applied to the modeling of the disjunction of concepts.
\end{abstract}

\section{Introduction}
Human thought makes use in a crucial way of concepts and how they are combined. The way knowledge is structured is also deeply related to concepts and their combinations. Hence fundamental questions about the nature of knowledge and human thought are related to the understanding of `how the meaning of a combination of concepts is related to the meaning of the concepts forming this combination'. Extensive studies of the problem of the combination of concepts have been undertaken over the years, but the problem is far from being solved \cite{hampton1991,hampton1997a,hampton1997b,kundamillerclaire1990,oshersonsmith1981,oshersonsmith1982,rips1995,smithosherson1984,smithoshersonripskeane1988,springermurphy1992}. Even for the most simple and basic combinations of concepts, such as conjunctions and disjunctions, the existing theories do not propose a satisfactory solution \cite{hampton1988a,hampton1988b,hampton1996,hampton1997a,oshersonsmith1981,rips1995,stormsdeboeckvanmechelengeeraerts1993,stormsdeboeckvanmechelenruts1996,stormsrutsvandenbroucke1998}.

Most research has been taking place on the problem of the conjunction of concepts. One of the first identifications of a problem situation -- now commonly referred to as the pet-fish problem, caused by the so-called guppy effect -- took place as a consequence of work by \cite{oshersonsmith1981}, by considering the concepts {\it Pet} and {\it Fish} and their conjunction {\it Pet-Fish}. Osherson and Smith observed that an item such as {\it Guppy} is considered to be very typical of the conjunction concept {\it Pet-Fish}, while not at all typical of both component concepts {\it Pet} and {\it Fish}. Hence the typicality of a specific item with respect to the conjunction of concepts can display unexpected behavior. The guppy effect was equally identified for the membership weight of items with respect to the conjunction of concepts \cite{hampton1988a,hampton1996,hampton1997a}.

To put this into perspective we mention that since the work of Eleanor Rosch and collaborators (see \cite{rosch1973a,rosch1973b,rosch1975,rosch1978,rosch1983}), cognitive scientists have commonly accepted that membership of an item for a specific concept category is in general not a `yes' or `no' notion, but a graded or fuzzy notion. This means that it can be characterized by means of a membership weight, which is a number between 0 and 1, where the value 1 of this number corresponds to membership of the concept category, the value 0 to non membership of the concept category, and values between 1 and 0 indicate a `graded or fuzzy amount of membership' of the item with respect to the considered concept. It is within this `graded membership' approach that \cite{hampton1988a} identified experimentally an effect similar to the guppy effect for typicality with respect to the conjunction of concepts, and Hampton named the effect `overextension'.

More concretely, for example in \cite{hampton1988a}, the concepts {\it Sports} and {\it Games} and their conjunction {\it Sports and Games} are considered. Next Hampton considers the item {\it Chess}, and it turns out that a test measuring how subjects estimate the membership weights of {\it Chess} for the concept {\it Sports} gives rise to the number 0.35, while the estimation of the membership weight for {\it Chess} for the concept {\it Games} gives rise to the number 0.94. When subjects are asked to estimate the membership weight of the item {\it Chess} for the combination {\it Sports and Games}, this gives rise to the number 0.79. This means that subjects find {\it Chess} to be `more strongly a member of the conjunction {\it Sports and Games}' than they find it to be a member of the concept {\it Sports} on its own. If one thinks intuitively of the `logical' meaning of a conjunction, this is a strange effect. Indeed, someone who finds that {\it Chess} is a {\it Sport and a Game}, would be expected to agree at least equally with the statement that {\it Chess} is a {\it Sport} if the conjunction of concept behaved in a similar way as the conjunction of logical propositions behaves.

The effect of overextension is abundant for the conjunction of concepts and has been studied intensively since \cite{oshersonsmith1981,oshersonsmith1982,smithosherson1984,smithoshersonripskeane1988,hampton1988a,hampton1997a,hampton1997b,stormsdeboeckvanmechelengeeraerts1993,stormsdeboeckvanmechelenruts1996,rips1995,stormsrutsvandenbroucke1998}. In \cite{hampton1988b} also the effect for the disjunction of concepts was studied, and a systematic underextension in this case was identified, with however also substantial cases of overextension. Both the experimental data for several extensive studies of the guppy effect for the conjunction of concepts, and the experimental data of Hampton's study of the underextension and overextension for the disjunction of concepts inspire the theory that we propose in this article.

We mentioned that membership of items of a concept is graded or fuzzy and mathematically represented by a membership weight. In this sense it is possible to frame such a description within the fuzzy set mathematics as introduced by \cite{zadeh1965}. Already \cite{oshersonsmith1981} showed that Zadeh's combination rules for the conjunction of fuzzy sets do not provide a good model for situations like the {\it Pet-Fish} one, and hence the guppy effect cannot be modeled well within fuzzy set theory. Also \cite{hampton1988b} showed that the underextension and overextension effects measured in the case of the disjunction of concepts cannot be accounted for well by means of a fuzzy set description of the disjunction. Before we give an exact description of how the theory we propose in this article does more than existing theories, let us look in some more detail at the case of the disjunction of concepts as studied in Hampton.

The experiment performed in \cite{hampton1988b} proceeds as follows. The tested subjects were a total of 40 undergraduate students. They were tested individually and were asked to choose a number from the following set: $\{-3,-2,-1,$ $0,+1,+2,+3\}$, with the positive numbers +1, +2 or +3 meaning that they considered `the item to be a member of the concept' and the typicality of the membership increasing with an increasing number. Hence the choice +3 means that the subject who attributes this number considers the item to be a very typical member, while the choice +1 means that he or she considers the item to be a not so typical member. The negative numbers indicate non membership, again in increasing order, hence -3 means strong non membership, while -1 means weak non membership. A choice of 0 stands for the subject's indecisiveness about the membership or non-membership of the item. Hampton used the list of concepts and their disjunctions shown in Table 1 below, and considered 24 items for each pair of concepts and their disjunction. Part of the items used by Hampton -- the ones showing the strongest effect of underextension or overextension -- are shown in Table 1.

\begin{table} 
\scriptsize
\begin{center}
{\begin{tabular}{lllll}
\hline
 & $\mu(A)$ & $\mu(B)$ & $\mu(A{\rm{\ or\ }}B)$ & $\Delta$ \\
 \hline
\multicolumn{5}{l}{\it $A$=House Furnishings, $B$=Furniture} \\
\hline
{\it Ashtray} & 0.7 & 0.3 & 0.25 & 0.45  \\
{\it Refrigerator} & 0.9 & 0.7 & 0.575 & 0.325 \\
{\it Park Bench} & 0 & 0.3 & 0.05 & 0.25 \\
{\it Waste-Paper Basket} & 1 & 0.5 & 0.6 & 0.4 \\
{\it Sink Unit} & 0.9 & 0.6 & 0.6 & 0.3 \\
\hline
\multicolumn{5}{l}{\it $A$=Hobbies, $B$=Games} \\
\hline
{\it Judo} & 1 & 0.7 & 0.8 & 0.2  \\
{\it Discus Throwing} & 1 & 0.75 & 0.7 & 0.3  \\
{\it Karate} & 1 & 0.7 & 0.8 & 0.2  \\
{\it Beer Drinking} & 0.8 & 0.2 & 0.575 & 0.225  \\
{\it Wrestling} & 0.9 & 0.6 & 0.625 & 0.275  \\
\hline
\multicolumn{5}{l}{\it $A$=Spices, $B$=Herbs} \\
\hline
{\it Monosodium Glutamate} & 0.15 & 0.1 & 0.425 & -0.275  \\
{\it Sugar} & 0 & 0 & 0.2 & -0.2  \\
{\it Sesame Seeds} & 0.35 & 0.4 & 0.625 & -0.225 \\
{\it Horseradish} & 0.2 & 0.4 & 0.7 & -0.3  \\
{\it Vanilla} & 0.6 & 0 & 0.275 & 0.325  \\
\hline
\multicolumn{5}{l}{\it $A$=Instruments, $B$=Tool} \\
\hline
{\it Pencil Eraser} & 0.4 & 0.7 & 0.45 & 0.25 \\
{\it Bicycle Pump} & 1 & 0.9 & 0.7 & 0.3  \\
{\it Computer} & 0.6 & 0.8 & 0.6 & 0.2  \\
{\it Spoon} & 0.65 & 0.9 & 0.7 & 0.2  \\
\hline
\multicolumn{5}{l}{\it $A$=Pets, $B$=Farmyard Animals} \\
\hline
{\it Camel} & 0.4 & 0 & 0.1 & 0.2  \\
{\it Guide Dog for the Blind} & 0.7 & 0 & 0.9 & -0.2  \\
{\it Spider} & 0.5 & 0.35 & 0.55 & -0.05  \\
{\it Monkey} & 0.5 & 0 & 0.25 & 0.25  \\
{\it Rat} & 0.5 & 0.7 & 0.4 & 0.3 \\
{\it Field Mouse} & 0.1 & 0.7 & 0.4 & 0.3 \\
\hline 
\multicolumn{5}{l}{\it $A$=Sportswear, $B$=Sports Equipment} \\
\hline
{\it Sunglasses} & 0.4 & 0.2 & 0.1 & 0.3  \\
{\it Lineman's Flag} & 0.1 & 1 & 0.75 & 0.25  \\
{\it Bathing Costume} & 1 & 0.8 & 0.8 & 0.2  \\
\hline
\multicolumn{5}{l}{\it $A$=Fruits, $B$=Vegetables} \\
\hline
{\it Parsley} & 0 & 0.2 & 0.45 & -0.25  \\
{\it Olive} & 0.5 & 0.1 & 0.8 & -0.3  \\
{\it Broccoli} & 0 & 0.8 & 1 & -0.2  \\
{\it Root Ginger} & 0 & 0.3 & 0.55 & -0.25 \\
{\it Tomato} & 0.7 & 0.7 & 1 & -0.3  \\
{\it Coconut} & 0.7 & 0 & 1 & -0.3  \\
{\it Mushroom} & 0 & 0.5 & 0.9 & -0.4 \\
{\it Green Pepper} & 0.3 & 0.6 & 0.8 & -0.2 \\
{\it Watercress} & 0 & 0.6 & 0.8 & -0.2  \\
{\it Garlic} & 0.1 & 0.2 & 0.5 & -0.3  \\
{\it Yam} & 0.45 & 0.65 & 0.85 & -0.2 \\
{\it Elderberry} & 1 & 0 & 0.8 & 0.2  \\
{\it Almond} & 0.2 & 0.1 & 0.425 & -0.225 \\
\hline
\multicolumn{5}{l}{\it $A$=Household Appliances, $B$=Kitchen Utensils} \\
\hline
{\it Cake Tin} & 0.4 & 0.7 & 0.95 & -0.25  \\
{\it Rubbish Bin} & 0.5 & 0.5 & 0.8 & -0.3  \\
{\it Electric Toothbrush} & 0.8 & 0 & 0.55 & 0.25  \\
\hline
\end{tabular}}
\end{center}
\caption{Membership weights and the `disjunction maximum rule deviation' of concepts and items of experiment 2 of \cite{hampton1988b}.}
\end{table}

\normalsize
The subjects were asked to repeat the procedure for all of the considered items and concepts. By means of the recollected data, membership weights were calculated by dividing the amount of positive ratings by the amount of non-zero ratings. For the item {\it Ashtray}, this led to 0.7 as membership weight for the concept {\it House Furnishings} and 0.3 as membership weight for the concept {\it Furniture}. For the disjunction {\it House Furnishings or Furniture}, a membership weight of 0.25 was measured. This means that subjects classified {\it Ashtray} to be less strongly a member of the concept {\it House Furnishings or Furniture} than of both concepts {\it House Furnishings} and {\it Furniture} apart, a deviation called `underextension' by \cite{hampton1988b}. The effect is abundant as one can see in Table 1, where for many items the membership weight of the disjunction is less than the membership weights of at least one of the concepts, and many times even less than the membership weights of both concepts.

Also the opposite effect occurred in the experiment. For example, for the concepts {\it Fruits}, {\it Vegetables} and {\it Fruits or Vegetables}, many items showed an opposite effect for the membership weights of the disjunction. Let us consider a specific example, namely the item {\it Olive}. In \cite{hampton1988b} the corresponding membership weights were measured for {\it Olive}, giving 0.5 as membership weight for {\it Fruits}, 0.1 as membership weight for {\it Vegetables}, and 0.8 as membership weight for {\it Fruits or Vegetables} was measured (see Table 1). Hence, subjects classified {\it Olive} to be a stronger member of the disjunction {\it Fruits or Vegetables} than it is a member of each of the concepts {\it Fruits} or {\it Vegetables} apart. This is a similar `overextension' to that encountered abundantly for the conjunction. Overextension could be fitted in a standard classical interpretation of the disjunction more easily than the more frequently occurring underextension, but not if the overextension is of the size as the one measured with some of the items, for example {\it Olive}. The data corresponding to the item {\it Olive}, but also many others in Table 1, are of this non classical overextended type.

In \cite{hampton1988b} different aspects of the deviation from the standard classical ways of interpreting the disjunction were analyzed coming to the conclusion that these standard classical approaches fail strongly in accounting for the measured data. In this article we show that the measured data can be modeled to a very satisfactory degree using the theory we propose.  This theory uses the mathematical apparatus of quantum mechanics, and the effects of overextension and underextension correspond to quantum interference effects and quantum context effects. Before embarking on the development of this quantum mechanical theory for cognition, let us be somewhat more specific on the classical analysis of the conjunction and disjunction and its deviations referred to as overextension and underextension.

If we describe the membership of an item for a concept by means of membership weights to take into account explicitly the graded and fuzzy structure of the membership notion, then from a mathematical point of view, we can still represent a concept by a set, if this set is a fuzzy set in the sense introduced in \cite{zadeh1965}. In fuzzy set theory the rules for conjunction and disjunction are the minimum and the maximum rule. More concretely, following these rules, the membership weight in the conjunction of two concepts equals the smallest of the two membership weights of the component concepts, and the membership weight of the disjunction is the greatest of the two membership weights of the component concepts. In \cite{oshersonsmith1981} it is shown how the situation of the pet-fish problem conflicts with the minimum rule of fuzzy set theory. In Table 1 we can see how individual items deviate from the maximum rule. We introduce
\begin{equation}
\Delta=\max (\mu(A),\mu(B)) - \mu(A{\rm{\ or\ }}B)
\end{equation}
which is the `maximum weight of both concepts minus the observed weight of the disjunction concept'. There are four disjunctions of concepts, which only show items with underextension (positive value of 
$\Delta$): {\it House Furnishings or Furniture}, {\it Hobbies or Games}, {\it Instruments or Tools} and {\it Sportswear or Sports Equipments}. The other disjunctions show items with underextension as well as overextension. One disjunction, {\it Pets or Farmyard Animals}, shows predominantly underextended items, while the remaining three disjunctions, {\it Spices or Herbs}, {\it Household Appliances or Kitchen Utensils} and {\it Fruits or Vegetables}. show predominantly overextended items.

In the next section we propose a quantum mechanical model for the situation, which produces, to a very great extent, the experimental data of \cite{hampton1988b}. We also show that it is three basic effects of quantum mechanics, interference, superposition and the effect of context which give rise to deviations from the classical situation that are exactly the same as Hampton's.

\section{Quantum Modeling of the Disjunction of Concepts} \label{feynman}

We introduce the aspects of quantum mechanics which we need step by step while we work out our general quantum modeling scheme.

\subsection{Hilbert Space, Kets, Bras and Complex Numbers}

In quantum mechanics a state of a quantum entity is described by a vector of length equal to 1. The Hilbert space of quantum mechanics is essentially the set of these vectors, hence each vector representing the state of the quantum entity under consideration, equipped with some additional structure. We denote vectors using the bra-ket notation introduced by Paul Adrien Dirac, one of the founding fathers of quantum mechanics (see \cite{dirac1958}), i.e.
$\left| A \right\rangle $,
$\left| B \right\rangle $. Vectors denoted in this way are called `kets', to distinguish them from another type of vectors, denoted as $\left\langle A \right|$,
$\left\langle B \right|$ and called `bras', which we will introduce later when we need them.

A state of a quantum entity is described by a ket vector, and by analogy we will describe the state of a concept by a ket vector. More concretely, consider the concept $A$, then the state of concept $A$ is represented by ket vector $\left| A \right\rangle$. We introduced the notion of `state of a concept' in detail in \cite{aertsgabora2005a}, and this is also the way we use it in the present article. The `state of a concept' represents `what the concept stands for with respect to its relevant features and contexts'.

The additional structure of a Hilbert space, as compared to being a vector space, is meant to express the notions of length, orthogonality and weight. This is achieved by introducing a product between a bra vector, for example  $\langle A|$, and a ket vector, for example $|B\rangle $, denoted as $\langle A|B\rangle$ and called a bra-ket. A bra-ket is always a complex number $re^{i\phi}$, which for historical reasons is called amplitude. We denote these complex numbers or amplitudes as $re^{i\phi}$, were $r$ is called the `magnitude' and $\phi$ the `phase' of the complex number. For those not acquainted with complex numbers, we make a short digression. The digression on the complex numbers will also be relevant to our subsequent attempt to understand the meaning of quantum interference and superposition in cognition.

Complex numbers were introduced by mathematicians with the original aim to be able to find numerical solutions to certain algebraic equations, which did not have solutions if only real numbers were taken into account. A simple example of such an equation is $x^2 = -1$. Indeed, the square of a real number is always positive, whether the number itself is positive or negative. This means that the equation $x^2 = -1$ has no solution if only real numbers are taken into account. Hence, mathematicians introduced a specific number, called the imaginary unit, denoted $i$, such that $i^2 = -1$. This number, together with all combinations (sums and products) of this number with the real numbers, yields the set of complex numbers. This means that an arbitrary complex number $z$  is of the form
\begin{equation}
z = x + iy
\end{equation}
where $x$ and $y$ are real numbers, and $i$ is the imaginary unit. We call $x$ the real part of the complex number $z$ and denote it by $\Re(z)$, and $y$ the imaginary part and denote it by $\Im(z)$.

It is easy to understand the outcomes of the sum and product of two complex numbers. Indeed, consider two complex numbers $z = x + iy$  and $z' = x' + iy'$. Let us calculate the sum $z + z'$
\begin{equation}
z + z' = (x + iy) + (x' + iy') = (x + x') + i(y + y')
\end{equation}
Hence, summing two complex numbers consists in summing the real parts and summing the imaginary parts of each individual complex number. The product of two complex numbers $z = x + iy$ and $z' = x' + iy'$ can also be calculated, taking into account that $i^2 = -1$. We have
\begin{equation}
zz' = xx' + xy'i + yx'i + yy'i^2  = (xx' - yy') + i(xy' + yx')
\end{equation}
It is important to note, because this is at the origin of the interference effect in quantum mechanics, that the imaginary parts of the complex numbers $z$  and $z'$  contribute to the real part of the product complex number $zz'$  with the factor $-yy'$. Hence some `real effect' comes from the `imaginary parts' if `products are made', and this is the root of the interference effect.

For each complex number $z = x + iy$  we introduce the conjugate complex number $z^* = x - iy$. By means of this conjugate number we can calculate the real part and imaginary part of the complex number. Indeed, for a complex number $z = x + iy$ we have
\begin{equation} \label{eq2.4}
\Re (z) = \frac{1}{2}(z + z*) = x \quad {\rm and} \quad \Im (z) = \frac{1}{{2i}}(z - z*) = y
\end{equation}
It is interesting to calculate $zz^*$
\begin{equation}
zz*=(x + iy)(x - iy)=x^2  - ixy + iyx - i^2 y^2=x^2  + y^2  
\end{equation}
This is a positive real number. The square root of this number, hence $\sqrt{x^2 + y^2}$, denoted $\left|z\right|$ or $r$, is called the `magnitude' of the complex numbers $z$ (and $z^*$, remark that both complex numbers $z$ and $z^*$ have the same magnitude).
\begin{equation}
r=\left| z \right| = \left| z^* \right| = \sqrt {x^2  + y^2 } 
\end{equation}
To get further into how complex numbers are commonly used in quantum mechanics, we introduce one of the fundamental formulas of mathematics, called Euler's formula. This formula gives a relation between the trigonometric functions of sine and cosine, the number $e$, which is the base number of the natural logarithm with value $e = 2.71828182 \dots$, and the imaginary unit $i$. The formula states that for an arbitrary angle $\phi$ we have
\begin{equation} \label{eq2.8}
e^{i\phi}  = \cos\phi  + i\sin\phi 
\end{equation}
This formula makes it possible to write an arbitrary complex number in a specific form, which is called the polar decomposition. Indeed, consider an arbitrary complex number $z$. We have
\begin{equation}
z = x + iy = r(\frac{x}{{\left| z \right|}} + i\frac{y}{{\left| z \right|}}) = r(\cos\phi  + i\sin \phi)  = re^{i\phi}
\end{equation}
where we have used that there exists an angle $\phi$ such that
\begin{equation}
{x \over r}=\cos\phi \quad {\rm and} \quad {y \over r}=\sin\phi 
\end{equation}
We can do this since
${x \over r}$ and ${y \over r}$ are real numbers smaller or equal to 1, such that the sum of their squares equals 1. The angle $\phi$ is called the `phase' of the complex number. The form
$z=re^{i\phi}$ is called the polar decomposition of the complex number $z$, and it is this form which is most commonly used in quantum mechanics. Remark that two complex numbers, written in their polar decomposition form $re^{i\phi}$ and $se^{i\sigma}$ are multiplied by multiplying their magnitudes and summing their phases, indeed we have
\begin{equation}
re^{i\phi} \cdot se^{i\sigma}=rs e^{i(\phi+\sigma)}
\end{equation}
Let us return to the elaboration of our quantum mechanical model to be used for the disjunction of concepts, and more specifically to a further explanation of the structure of Hilbert space.

The absolute value of the complex number  $\langle A|B\rangle$ is equal to the length of  $|A\rangle$ times the length of  $|B\rangle$ times the cosine of the angle between vectors  $|A\rangle$ and  $|B\rangle$. From this it follows that we have a definition of the length of a ket and bra vector
\begin{equation}
\||A\rangle\|=\|\langle A|\|=\sqrt{\langle A|A\rangle}
\end{equation}
We mentioned already that in quantum mechanics a state of the quantum entity is represented by means of a ket vector of length 1. Hence we can now specify this requirement for the vectors involved with concepts $A$ and $B$. Vectors $|A\rangle$ and $|B\rangle$ are such that
\begin{equation}
1=\langle A|A\rangle=\langle B|B\rangle
\end{equation}
We mentioned that $\langle A|B\rangle$
 is a complex number of which the absolute value equals the length of $|A\rangle$ times the length of $|B\rangle$ times the cosine of the angle between $|A\rangle$ and $|B\rangle$. This means that $|A\rangle$  and $|B\rangle$  are orthogonal, in the sense that the angle between both vectors is $90^\circ$, if  $\langle A|B\rangle= 0$. We denote this as follows $|A\rangle \perp |B\rangle$. 

We mentioned that $\langle A|B\rangle$
  is a complex number; additionally, in the quantum formalism $\langle A|B\rangle$
  is the complex conjugate of $\langle B|A\rangle$. Hence
\begin{equation}
\langle B|A\rangle^*=\langle A|B\rangle
\end{equation}
Further, the operation bra-ket $\langle \cdot | \cdot \rangle$ is linear in the ket and anti-linear in the bra. Hence
\begin{eqnarray} \label{eq3.5a}
&&\langle A|(x|B\rangle+y|C\rangle)=x\langle A|B\rangle+y\langle A|C\rangle \\ \label{eq3.5b}
&&(a\langle A|+b\langle B|)|C\rangle=a^*\langle A|C\rangle+b^*\langle B|C\rangle
\end{eqnarray}
In our digression on complex numbers we saw that the absolute value of a complex number is defined as the square root of the product of this complex number and its complex conjugate. Hence we have
\begin{equation}
|\langle A|B\rangle|=\sqrt{\langle A|B\rangle \langle B|A\rangle}
\end{equation}
An orthogonal projection $M$ is a linear function on the Hilbertspace, hence $M: {\cal H} \rightarrow {\cal H} |A\rangle \mapsto M|A\rangle$, which is hermitean and idempotent, which means that for $|A\rangle, |B\rangle \in {\cal H}$ and $x, y \in \compl$ we have
\begin{eqnarray} \label{linearity}
&M(z|A\rangle+t|B\rangle)=zM|A\rangle+tM|B\rangle \quad {\rm (linearity)} \\ \label{hermiticity}
&\langle A|M|B\rangle=\langle B|M|A\rangle \quad {\rm (hermiticity)} \\ \label{idempotenty}
&M \cdot M=M \quad {\rm(idempotenty)}
\end{eqnarray}
Measurable quantities, often called observables in quantum mechanics, are represented by means of hermitian linear functions on the Hilbert space, and for two valued observables these Hermitian functions are orthogonal projections. This is the reason that we can describe the decision measurement of `being member of' or `not being member of' with respect to a concept by means of an orthogonal projection on the Hilbert space. Concretely, let us consider an item $X$, then the decision measurement of `the item $X$ being a member of the concept $A$' is represented by means of the orthogonal projection $M$.

If in quantum mechanics an orthogonal projection $M$ represents a yes-no measurement of a certain observable, then the quantum equation determining the probability $\mu(A)$ for outcome `yes' if this measurement is executed on a quantum entity in state $|A\rangle$ is given by $\langle A|M|A\rangle$. Hence, in the case of the measurement being the `yes, no' decision for membership of item $X$ with respect to concept $A$, we have that the probability $\mu(A)$ for a subject to take decision `yes', in favor of membership, is given by 
\begin{equation}
\mu(A)=\langle A|M|A\rangle
\end{equation}

\subsection{The Quantum Effect of Interference} \label{interference}
From now on we consider the situation of two concepts $A$ and $B$, and an item $X$ such that the membership of item $X$ with respect to both concepts $A$ and $B$ and also with respect to the disjunction `$A$ or $B$' of both concepts, can be considered, and we denote these membership weights  by $\mu(A)$, $\mu(B)$ and $\mu(A\ {\rm or}\ B)$, respectively.

Concepts $A$ and $B$ are descibed by vectors $|A\rangle, |B\rangle \in {\cal H}$ and for the item $X$, the observable `$X$ is member of' is described by the orthogonal projection $M: {\cal H} \rightarrow {\cal H}$, and hence we have
\begin{equation}
\mu(A)=\langle A|M|A\rangle \quad \mu(B)=\langle B|M|B\rangle
\end{equation}
This orthogonal projection $M$ determines two orthogonal subspaces $M({\cal H})$ and $(1-M)({\cal H})$ of the Hilbert space ${\cal H}$. The vectors $M|A\rangle$ and $M|B\rangle$ are the projections of the vectors $|A\rangle$ and  $|B\rangle$ on the subspace $M({\cal H})$, respectively, and the vectors $(1-M)|A\rangle$ and $(1-M)|B\rangle$ are the projections of the vector $|A\rangle$ and $|B\rangle$ on the subspace $(1-M)({\cal H})$, respectively, and they are orthogonal vectors, indeed
\begin{eqnarray}
\langle A|M(1-M)|A\rangle&=&\langle A|M|A\rangle-\langle A|M^2|A\rangle \\
&=&\langle A|M|A\rangle-\langle A|M|A\rangle=0 \\
\langle B|M(1-M)|B\rangle&=&\langle B|M|B\rangle-\langle B|M^2|B\rangle \\
&=&\langle B|M|B\rangle-\langle B|M|B\rangle=0
\end{eqnarray}
where we make explicit use of the idempotenty property of an orthogonal projection, i.e. equation (\ref{idempotenty}), or $M^2=M$, and which shows that $M|A\rangle \perp (1-M)|A\rangle$ and $M|B\rangle \perp (1-M)|B\rangle$. Let us call $|e\rangle$ a unit vector on the ray determined by vector $M|A\rangle$, $|e'\rangle$ a unit vector on on the ray determined by vector $(1-M)|A\rangle$, $|f\rangle$ a unit vector on on the ray determined by vector $M|B\rangle$ and $|f'\rangle$ a unit vector on on the ray determined by vector $(1-M)|B\rangle$. There then exists $a, \alpha, a', \alpha', b, \beta, b'$ and $\beta'$ such that
\begin{equation} \label{formAandformB}
|A\rangle=ae^{i\alpha}|e\rangle+a'e^{i\alpha'}|e'\rangle \quad {\rm and} \quad |B\rangle=be^{i\beta}|f\rangle+b'e^{i\beta'}|f'\rangle
\end{equation}
and we have $\langle e|e\rangle=\langle e'|e'\rangle=\langle f|f\rangle=\langle f'|f'\rangle=1$ and $\langle e|e'\rangle=\langle e'|e\rangle=\langle f|f'\rangle=\langle f'|f\rangle=0$.
Before proceeding, let us calculate $\langle A|B \rangle$ if we introduce explicit forms for the complex numbers $ \langle e|f \rangle $ and $ \langle e'|f' \rangle $, namely
\begin{equation}
\langle e|f\rangle=ce^{i\gamma} \quad \langle e'|f'\rangle=c'e^{i\gamma'}
\end{equation}
and angles
\begin{equation} \label{phiandphi'}
\phi= \beta - \alpha + \gamma \quad \phi'= \beta' - \alpha' + \gamma' 
\end{equation}
We have
\begin{eqnarray}
\langle A|B \rangle &=& (ae^{-i\alpha} \langle e|+a'e^{-i\alpha'}\langle e'|)(be^{i\beta}|f\rangle+b'e^{i\beta'}|f'\rangle)  \\
&=&abe^{i(\beta-\alpha)}\langle e|f \rangle + a'b'e^{i( \beta'- \alpha')}\langle e'|f' \rangle  \\ \label{AinpruductB}
&=&abce^{i(\beta-\alpha+\gamma)}+a'b'c'e^{i(\beta'-\alpha'+\gamma')}=abce^{i\phi}+a'b'c'e^{i\phi'}
\end{eqnarray} 
Further we have
\begin{eqnarray}
\mu(A)&=&\langle A|M|A\rangle
=(ae^{-i\alpha}\langle e|+a'e^{-i\alpha'}\langle e'|)(ae^{i\alpha}|e\rangle)  \\
&=&a^2 \langle e|e\rangle+aa'e^{i(\alpha'-\alpha)}\langle e'|e\rangle =a^2 \\
\mu(B)&=&\langle B|M|B\rangle=(be^{-i\beta}\langle f|+b'e^{-i\beta'}\langle f'|)(be^{i\beta}|f\rangle)  \\
&=&b^2 \langle f|f\rangle+bb'e^{i(\beta'-\beta)}\langle f'|f\rangle=b^2
\end{eqnarray}
We have not yet put forward how we will describe quantum mechanically the concept `$A$ or $B$'. The idea is to use the procedure in which the `or' of two situations is described in quantum mechanics, namely by means of the quantum superposition of the states that describe the individual situations apart. Think for example of the archetypical double slit experiment situation, where the situation of `both slits open' is described by the quantum superposition of the two states of which each one describes the situation of one of slits open and the other one closed. 

Hence, the concept `$A$ or $B$' is represented by means of the `normalized superposition' of the vectors that represent concepts $A$ and $B$. Let us suppose first that these vectors are orthogonal, hence $|A\rangle \perp |B\rangle$, which means that $\langle A|B\rangle=\langle B|A\rangle=0$. Let us consider this normalized superposition
\begin{equation}
|AB\rangle={1 \over \sqrt{2}}(|A\rangle+|B\rangle) 
\end{equation}
Then we have
\begin{eqnarray}
\langle AB|AB\rangle&=&{1 \over 2}(\langle A|+\langle B|)(|A\rangle+|B\rangle)  \\
&=&{1 \over 2}(\langle A|A\rangle+\langle A|B\rangle+\langle B|A\rangle+\langle B|A\rangle) = 1
\end{eqnarray}
which shows that $|AB\rangle$ is a unit vector. We can now calculate the probabilities $\mu(A\ {\rm or}\ B)$ for membership of the item $X$ with respect to the disjunction concept `$A$ or $B$'.
\begin{eqnarray}
\mu(A\ {\rm or}\ B)&=&{1 \over 2}(\langle A|+\langle B|)M(|A\rangle+|B\rangle)  \\
&=&{1 \over 2}(\langle A|M|A\rangle+\langle A|M|B\rangle+\langle B|M|A\rangle+\langle B|M|B\rangle)  \\
&=&{1 \over 2}(\mu(A)+\langle A|M|B\rangle+\langle A|M|B\rangle^*+\mu(B))  \\
&=&{\mu(A)+\mu(B) \over 2}+\Re\langle A|M|B\rangle\end{eqnarray}
Let us calculate $\langle A|M|B\rangle$ making use of (\ref{formAandformB}). We have
\begin{eqnarray}
\langle A|M|B\rangle&=&(ae^{-i\alpha}\langle e|+a'e^{-i\alpha'}\langle e'|)M(be^{i\beta}|f\rangle+b'e^{i\beta'}|f'\rangle) \\
&=&ae^{-i\alpha}\langle e|M|be^{i\beta}|f\rangle \\
&=&abe^{i(\beta-\alpha)}\langle e|f\rangle=c\sqrt{\mu(A)\mu(B)}e^{i\phi}
\end{eqnarray}
Hence this gives
\begin{equation} \label{orthosolution}
\mu(A\ {\rm or}\ B)={\mu(A)+\mu(B) \over 2}+c\sqrt{\mu(A)\mu(B)}\cos\phi
\end{equation}
We have not yet calculated the relation between $a, b', c'$ and $a, b, c$ due to $|A \rangle \perp |B \rangle $. Let us first suppose that $a'b'c'\not=0$. Then from (\ref{AinpruductB}) follows that
\begin{eqnarray}
&&0=abce^{i\phi}+a'b'c'e^{i\phi'} \Leftrightarrow 0=abc+a'b'c'e^{i(\phi'-\phi)} \\
&\Leftrightarrow&0=abc+a'b'c'\cos(\phi'-\phi)+ia'b'c'\sin(\phi'-\phi) \\
&\Leftrightarrow&0=abc+a'b'c'\cos(\phi'-\phi) \quad {\rm and} \quad  0=a'b'c'\sin(\phi'-\phi) \\
&\Leftrightarrow&0=abc+a'b'c'\cos(\phi'-\phi)\quad {\rm and} \quad 0=\sin(\phi'-\phi) \\
&\Leftrightarrow&0=abc+a'b'c'\cos(\phi'-\phi)\quad {\rm and} \quad \phi'-\phi=k\pi \\
&\Leftrightarrow& abc=a'b'c'\quad {\rm and} \quad  \phi'-\phi=(2k+1)\pi
\end{eqnarray}
We find an equivalent condition if $abc\not=0$. Hence this means that we have $abc=a'b'c'=0$ and in this case $\phi$ and $\phi'$ are arbitrary, or we have $abc\not=0$ or $a'b'c'\not=0$ and then we have $abc=a'b'c'$ and $\phi'-\phi=(2k+1)\pi$. This is equivalent to 
\begin{equation}
abc=a'b'c'
\end{equation}
in any case, and 
\begin{equation} \label{ortho01}
\phi-\phi'=(2k+1)\pi \quad {\rm if} \quad abc\not=0
\end{equation}
From this follows that
\begin{eqnarray} \label{ortho02}
1-\mu(A\ {\rm or}\ B)&=&{2-\mu(A)-\mu(B) \over 2}+c'\sqrt{(1-\mu(A))(1-\mu(B)}\cos\phi' \\
&=&{2-\mu(A)-\mu(B) \over 2}-c\sqrt{\mu(A)\mu(B)}\cos\phi
\end{eqnarray} 
We construct an explicit solution that allows the greatest possible interference in the following way. In case $\mu(A)\mu(B) \le (1-\mu(A))(1-\mu(B))$ we choose $c=1$ and $c'$ such that $a'b'c'=ab$, and in case $(1-\mu(A))(1-\mu(B)) < \mu(A)\mu(B)$ we choose $c'=1$ and $c$ such that $abc=a'b'$. With these choices we get
\begin{eqnarray} \label{orthosolution}
\mu(A\ {\rm or}\ B)&=&{\mu(A)+\mu(B) \over 2} \nonumber \\
&&+\sqrt{\min(\mu(A)\mu(B),(1-\mu(A))(1-\mu(B))}\cos\phi
\end{eqnarray}
We can now also easily see for which data a quantum model exists. Indeed, the minimum value and maximum value of (\ref{orthosolution}) is reached for $\phi=\pi$ and for $\phi=0$, respectively. This means that a quantum model exists if $\mu(A\ {\rm or}\ B)$ is contained in the interval $I_{orth}$ starting at this minimum and ending at this maximum
\begin{eqnarray}
I_{orth}&=&[{\mu(A)+\mu(B) \over 2}-\sqrt{\min(\mu(A)\mu(B),(1-\mu(A))(1-\mu(B))}, \nonumber \\
&&{\mu(A)+\mu(B) \over 2}+\sqrt{\min(\mu(A)\mu(B),(1-\mu(A))(1-\mu(B))}]
\end{eqnarray}
We call $I_{orth}$ the `orthogonal solution interval' and in Table 2 we represent $I_{orth}$ for the items of Table 1 tested in \cite{hampton1988b}.

\subsection{Construction of a $\compl^3$ Representation}

Let us make the Hilbert space representation more concrete. First we remark that for arbitrary $\mu(A)$ and $\mu(B)$ we always have one of the two possibilities, (i) $\mu(A)+\mu(B)\le1$ or (ii) $1<\mu(A)+\mu(B)$, and we will built a slightly different model for both cases.

(i) Consider first the case where $\mu(A)+\mu(B)\le1$. This means that $\mu(A)\mu(B) \le (1-\mu(A))(1-\mu(B))$. Consider $\compl^3$ and choose for subspace $M({\cal H})$ the ray generated by $(0,0,1)$, and for subspace $(1-M)({\cal H})$ the plane generated by $(1,0,0)$ and $(0,1,0)$. For vectors $|A\rangle$ and $|B\rangle$ and phase $\phi$ we choose

\begin{table}
\scriptsize
\begin{center}
{\begin{tabular}{lll}
\hline 
 & $\mu(A{\rm{\ or\ }}B)$ & $I_{orth}$  \\ 
\hline  
\multicolumn{3}{l}{\it $A$=House Furnishings, $B$=Furniture} \\
\hline
{\it Ashtray} & 0.25 & [0.0417, 0.9583]  \\
{\it Refrigerator} & 0.575 & [0.6268, 0.9732] \\
{\it Park Bench} & 0.05 & [0.15, 0.15] \\
{\it Waste-Paper Basket} & 0.6 & [0.75, 0.75] \\
{\it Sink Unit} & 0.6 & [0.55, 0.95]  \\
\hline
\multicolumn{3}{l}{\it $A$=Hobbies, $B$=Games} \\
\hline
{\it Judo} & 0.8 & [0.85, 0.85] \\
{\it Discus Throwing} & 0.7 & [0.875, 0.875] \\
{\it Karate} & 0.8 & [0.85, 0.85] \\
{\it Beer Drinking} & 0.575 & [0.1, 0.9]  \\
{\it Wrestling} & 0.625 & [0.55, 0.95]   \\
\hline
\multicolumn{3}{l}{\it $A$=Spices, $B$=Herbs} \\
\hline
{\it Monosodium Glutamate} & 0.425 & [0.0025, 0.2475] \\
{\it Sugar} & 0.2 & [0, 0] \\
{\it Sesame Seeds} & 0.625 & [0.0008, 0.7492]  \\
{\it Horseradish} & 0.7 & [0.0172, 0.5828] \\
{\it Vanilla} & 0.275 & [0.3, 0.3] \\
\hline
\multicolumn{3}{l}{\it $A$=Instruments, $B$=Tool} \\
\hline
{\it Pencil Eraser} & 0.45 & [0.1257, 0.9743] \\
{\it Bicycle Pump} & 0.7 & [0.95, 0.95] \\
{\it Computer} & 0.6 & [0.4172, 0.9828] \\
{\it Spoon} & 0.7 & [0.5879, 0.9621] \\
\hline
\multicolumn{3}{l}{\it $A$=Pets, $B$=Farmyard Animals} \\
\hline
{\it Camel} & 0.1 & [0.2, 0.2] \\
{\it Guide Dog for the Blind} & 0.9 & [0.35, 0.35] \\
{\it Spider} & 0.55 & [0.0067, 0.8433]  \\
{\it Monkey} & 0.25 & [0.25, 0.25]  \\
{\it Rat} & 0.4 & [0.2127, 0.9873]  \\
{\it Field Mouse} & 0.4 & [0.1354, 0.6646]  \\
\hline 
\multicolumn{3}{l}{\it $A$=Sportswear, $B$=Sports Equipment} \\
\hline
{\it Sunglasses} & 0.1 & [0.0172, 0.5828]  \\
{\it Lineman's Flag} & 0.75 & [0.55, 0.55]  \\
{\it Bathing Costume} & 0.8 & [0.9, 0.9]  \\
\hline
\multicolumn{3}{l}{\it $A$=Fruits, $B$=Vegetables} \\
\hline
{\it Parsley} & 0.45 & [0.1, 0.1]  \\
{\it Olive} & 0.8 & [0.0764, 0.5236]  \\
{\it Broccoli} & 1 & [0.4, 0.4]  \\
{\it Root Ginger} & 0.55 & [0.15, 0.15]  \\
{\it Tomato} & 1 & [0.4, 1]  \\
{\it Coconut} & 1 & [0.35, 0.35]  \\
{\it Mushroom} & 0.9 & [0.25, 0.25]  \\
{\it Green Pepper} & 0.8 & [0.0257, 0.8743]  \\
{\it Watercress} & 0.8 & [0.3, 0.3]  \\
{\it Garlic} & 0.5 & [0.0086, 0.2914]  \\
{\it Yam} & 0.85 & [0.1113, 0.9887]  \\
{\it Elderberry} & 0.8 & [0.5, 0.5]  \\
{\it Almond} & 0.425 & [0.0086, 0.2914]  \\
\hline
\multicolumn{3}{l}{\it $A$=Household Appliances, $B$=Kitchen Utensils} \\
\hline
{\it Cake Tin} & 0.95 & [0.1257, 0.9743]  \\
{\it Rubbish Bin} & 0.8 & [0, 1]  \\
{\it Electric Toothbrush} & 0.55 & [0.4, 0.4]  \\
\hline
\end{tabular}}
\end{center}
\caption{The `orthogonal solution intervals' $I_{orth}$ for concepts and items of experiment 2 of \cite{hampton1988b}.}
\end{table}

\normalsize
\begin{equation} \label{vectorA01}
|A\rangle=(\sqrt{1-\mu(A)},0,\sqrt{\mu(A)}) 
\end{equation} 
\begin{eqnarray} \label{vectorB01}
|B\rangle&=&e^{i\phi}(-{\sqrt{\mu(A)\mu(B)} \over \sqrt{1-\mu(A)}},-\sqrt{{1-\mu(B)-\mu(A) \over 1-\mu(A)}},\sqrt{\mu(B)})  \\ 
&&\quad {\rm if} \ \mu(A)\not=1 \nonumber \\
&=&e^{i\phi}(0,1,0) \quad {\rm if} \quad \mu(A)=1 
\end{eqnarray}
\begin{eqnarray} \label{anglephi01}
\phi&=&\arccos({2\mu(A\ {\rm or}\ B)-\mu(A)-\mu(B) \over 2\sqrt{\mu(A)\mu(B)}}) \quad {\rm if} \quad \mu(A)\mu(B)\not=0 \\
&=&{\rm arbitrary} \quad {\rm if} \quad \mu(A)\mu(B)=0
\end{eqnarray} 
Let us verify that this is a solution. We have $\langle A|A\rangle=(1-\mu(A))+\mu(A)=1$. In case  $\mu(A)\not=1$, we have $\langle B|B\rangle={\mu(A)\mu(B) \over 1-\mu(A)}+{1-\mu(B)-\mu(A) \over 1-\mu(A)}+\mu(B)={(1-\mu(B))(1-\mu(A)) \over 1-\mu(A)}+\mu(B)=(1-\mu(B))+\mu(B)=1$, and also in case $\mu(A)=1$ we have $\langle B|B\rangle=1$. This shows that both vectors $|A\rangle$ and $|B\rangle$ are unit vectors. We have $\langle A|B\rangle=-e^{i\phi}\sqrt{1-\mu(A)}{\sqrt{\mu(A)\mu(B)} \over \sqrt{1-\mu(A)}}+e^{i\phi}\sqrt{\mu(A)\mu(B)}=0$ in case $\mu(A)\not=1$, and also $\langle A|B\rangle=0$ in case $\mu(A)=1$. This shows that $|A\rangle$ and $|B\rangle$ are orthogonal. In case $\mu(A)\not=1$ we have $\langle A|M|B\rangle=e^{i\phi}\sqrt{\mu(A)\mu(B)}$ and hence $\Re\langle A|M|B\rangle=\sqrt{\mu(A)\mu(B)}\cos\phi$. In case $\mu(A)\mu(B)\not=0$ we have $\cos\phi={2\mu(A\ {\rm or}\ B)-\mu(A)-\mu(B) \over 2\sqrt{\mu(A)\mu(B)}}$, and hence $\Re\langle A|M|B\rangle=\mu(A\ {\rm or}\ B)-{1 \over 2}(\mu(A)+\mu(B))$ from which follows that $\mu(A\ {\rm or}\ B)={1 \over 2}(\mu(A)+\mu(B))+\Re\langle A|M|B\rangle$. In case $\mu(A)\mu(B)=0$ we have $\Re\langle A|M|B\rangle=0$, and hence only when $\mu(A\ {\rm or}\ B)={1 \over 2}(\mu(A)+\mu(B))$ we have a solution.

(ii) Let us construct now a solution for the case where $1<\mu(A)+\mu(B)$. We then have $(1-\mu(A))(1-\mu(B)) < \mu(A)\mu(B)$. Again we consider $\compl^3$ but this time take for $M({\cal H})$ the subspace generated by the vectors $(1,0,0)$ and $(0,1,0)$, and for $(1-M)({\cal H})$ the ray generated by the vector $(0,0,1)$. For vectors $|A\rangle$, $|B\rangle$ and phase $\phi$ we choose
\begin{equation} \label{vectorA02}
|A\rangle=(\sqrt{\mu(A)},0,\sqrt{1-\mu(A)})
\end{equation}
\begin{eqnarray} \label{vectorB02}
|B\rangle&=&e^{i\phi}(\sqrt{(1-\mu(A))(1-\mu(B)) \over \mu(A)},\sqrt{\mu(A)+\mu(B)-1 \over \mu(A)},\nonumber\\
&&-\sqrt{1-\mu(B)}) \quad {\rm if} \quad \mu(A)\not=0 \\
&=&e^{i\phi}(0,1,0) \quad {\rm if} \quad \mu(A)=0 
\end{eqnarray}
\begin{eqnarray} \label{anglephi02}
\phi&=&\arccos({2\mu(A\ {\rm or}\ B)-\mu(A)-\mu(B) \over 2\sqrt{(1-\mu(A))(1-\mu(B))}}) \quad {\rm if} \quad \mu(A)\mu(B)\not=1 \\
&=&\quad {\rm arbitrary} \quad {\rm if} \quad \mu(A)\mu(B)=1
\end{eqnarray}
Let us also verify whether this is a solution. We have $\langle A|A\rangle=\mu(A)+(1-\mu(A))=1$. In case $\mu(A)\not=0$ we have $\langle B|B\rangle={(1-\mu(A))(1-\mu(B)) \over \mu(A)}+{\mu(A)+\mu(B)-1 \over \mu(A)}+1-\mu(B)={\mu(A)\mu(B) \over \mu(A)}+1-\mu(B)=1$, and also in case $\mu(A)=0$ we have $\langle B|B\rangle=1$. This shows that $|A\rangle$ and $|B\rangle$ are unit vectors. We have $\langle A|B\rangle=e^{i\phi}\sqrt{\mu(A)}\sqrt{(1-\mu(A))(1-\mu(B)) \over \mu(A)}-e^{i\phi}\sqrt{1-\mu(A)}\sqrt{1-\mu(B)}=0$ if $\mu(A)\not=0$, and also $\langle A|B\rangle=0$ if $\mu(A)=0$, which shows that $|A\rangle$ and $|B\rangle$ are orthogonal. In case $\mu(A)\not=0$ we have $\langle A|M|B\rangle=e^{i\phi}\sqrt{(1-\mu(A))(1-\mu(B))}$ and hence $\Re\langle A|M|B\rangle=\sqrt{(1-\mu(A))(1-\mu(B))}\cos\phi$. In case $\mu(A)\mu(B)\not=1$ we have $\cos\phi={2\mu(A\ {\rm or}\ B)-\mu(A)-\mu(B) \over 2\sqrt{(1-\mu(A))(1-\mu(B))}}$ and hence $\Re\langle A|M|B\rangle=\mu(A\ {\rm or}\ B)-{1 \over 2}(\mu(A)+\mu(B))$ from which follows that $\mu(A\ {\rm or}\ B)={1 \over 2}(\mu(A)+\mu(B))+\Re\langle A|M|B\rangle$. In case $\mu(A)\mu(B)=1$ we have $\Re\langle A|M|B\rangle=0$, and hence only when $\mu(A\ {\rm or}\ B)={1 \over 2}(\mu(A)+\mu(B))$ we have a solution.

Let us work out some examples. Consider the item {\it Pencil Eraser} with respect to the pair of concepts {\it Instruments} and {\it Tools} and their disjunction {\it Instruments or Tools}. In \cite{hampton1988b} the membership weights were measured giving $\mu(A)=0.4$, $\mu(B)=0.7$ and $\mu(A\ {\rm or}\ B)=0.45$. Let us construct the $\compl^3$ quantum model for this item. We have $\mu(A)+\mu(B)=1.1$, and hence making use of equations (\ref{vectorA02}), (\ref{vectorB02}) and (\ref{anglephi02}), we find $|A\rangle=(0.6325, 0, 0.7746)$, $|B\rangle=e^{i\phi}(0.6708, 0.5, -0.5477)$ and $\phi=103.6330^\circ$. As a second example we consider the item {\it Ashtray} with respect to the pair of concepts {\it House Furnishings} and {\it Furniture} and their disjunction {\it House Furnishings or Furniture}. In \cite{hampton1988b} the membership weights were measured giving $\mu(A)=0.7$, $\mu(B)=0.3$ and $\mu(A\ {\rm or}\ B)=0.25$. For the $\compl^3$ realization we find $|A\rangle=(0.5477, 0, 0.8367)$, $|B\rangle=e^{i\phi}(-0.8367, 0, 0.5477)$ and $\phi=123.0619^\circ$. Our final example concerns the item {\it Field Mouse} with respect to the pair of concepts {\it Pets} and {\it Farmyard Animals} and their disjunction {\it Pets or Farmyard Animals}. For this item, the values $\mu(A)=0.1$, $\mu(B)=0.7$ and $\mu(A\ {\rm or}\ B)=0.4$. For the $\compl^3$ realization we find $|A\rangle=(0.9487, 0, 0.3162)$, $|B\rangle=e^{i\phi}(-0.2789, -0.4714, 0.8367)$ and $\phi=90^\circ$.

In this $\compl^3$ model, only $e^{i\phi}$ appears as `not a real number' in the vector $|B\rangle$. For two values of $\phi$, namely $\phi=0^\circ$ and $\phi=180^\circ$, $e^{i\phi}$ is a real number. The interference effect is present for these two values of $\phi$ but can take only two values, and these values are the minimum and maximum amounts of interference . The role of the complex numbers is to allow it to obtain any value in between these two values. In Table 3 we have calculated the vectors $|A\rangle$ and $|B\rangle$ and the angle $\phi$ for the experimental data of \cite{hampton1988b} for which a solution exists. For those items where no vectors are shown in Table 3 it means that the $\compl^3$ model does not exist, i.e. when $\mu(A\ {\rm or}\ B)\notin I_{orth}$.

\begin{table}
\scriptsize
\begin{center}
{\begin{tabular}{lll}
\hline 
 & $|A\rangle$ & $|B\rangle$ \\ 
\hline  
\multicolumn{3}{l}{\it $A$=House Furnishings, $B$=Furniture} \\
\hline
{\it Ashtray} & (0.5477, 0, 0.8367) & $e^{i123.0619^\circ}$(-0.8367, 0, 0.5477) \\
{\it Sink Unit} & (0.9487, 0, 0.3162) & $e^{i138.5904^\circ}$(0.2108, 0.7454,-0.6325) \\
\hline
\multicolumn{3}{l}{\it $A$=Hobbies, $B$=Games} \\
\hline
{\it Beer Drinking} &  (0.4472, 0, 0.8944) & $e^{i79.1931^\circ}$(-0.8944, 0, 0.4472)  \\
{\it Wrestling} &  (0.9487, 0, 0.3162) & $e^{i128.6822^\circ}$(0.2108, 0.7454, -0.6325)   \\
\hline
\multicolumn{3}{l}{\it $A$=Spices, $B$=Herbs} \\
\hline
{\it Sesame Seeds} &  (0.8062, 0, 0.5916) & $e^{i48.0753^\circ}$(-0.4641, -0.6202, 0.6325) \\
\hline
\multicolumn{3}{l}{\it $A$=Instruments, $B$=Tool} \\
\hline
{\it Pencil Eraser} &  (0.6325, 0, 0.7746) & $e^{i103.6330^\circ}$(0.6708, 0.5, -0.5477) \\
{\it Computer} &  (0.7746, 0, 0.6325) & $e^{i110.7048^\circ}$(0.3651, 0.8165, -0.4472)\\
{\it Spoon} &  (0.8062, 0, 0.5916) & $e^{i113.6339^\circ}$(0.2320, 0.9199, -0.3162) \\
\hline
\multicolumn{3}{l}{\it $A$=Pets, $B$=Farmyard Animals} \\
\hline
{\it Spider} &  (0.7071, 0, 0.7071) & $e^{i72.6140^\circ}$(-0.5916, -0.5477, 0.5916) \\
{\it Monkey} &  (0.7071, 0, 0.7071) & $e^{iarbitrary^\circ}$(0, -1, 0) \\
{\it Rat} &   (0.7071, 0, 0.7071) & $e^{i121.0909^\circ}$(0.5477, 0.6325, -0.5477) \\
{\it Field Mouse} &  (0.9487, 0, 0.3162) & $e^{i90^\circ}$(-0.2789, -0.4714, 0.8367) \\
\hline 
\multicolumn{3}{l}{\it $A$=Sportswear, $B$=Sports Equipment} \\
\hline
{\it Sunglasses} &  (0.7746, 0, 0.6325) & $e^{i135^\circ}$(-0.3651, -0.8165, 0.4472) \\
\hline
\multicolumn{3}{l}{\it $A$=Fruits, $B$=Vegetables} \\
\hline
{\it Tomato} &  (0.8367, 0, 0.5477) & $e^{i0^\circ}$(0.3586, 0.7559, -0.5477) \\
{\it Green Pepper} &  (0.8367, 0, 0.5477) & $e^{i34.4158^\circ}$(-0.5071, -0.3780, 0.7746) \\
{\it Yam} &  (0.6708, 0, 0.7416) & $e^{i46.8616^\circ}$(0.6540, 0.4714, -0.5916) \\
\hline
\multicolumn{3}{l}{\it $A$=Household Appliances, $B$=Kitchen Utensils} \\
\hline
{\it Cake Tin} &  (0.6325, 0, 0.7746) & $e^{i19.4712^\circ}$(0.6708, 0.5, -0.5477) \\
{\it Rubbish Bin} &  (0.7071, 0, 0.7071) & $e^{i53.1301^\circ}$(-0.7071, 0, 0.7071) \\
\hline
\end{tabular}}
\end{center}
\caption{`Orthogonal solutions' for concepts and items of experiment 2 of \cite{hampton1988b}.}
\end{table}

\normalsize
\section{The Joint Context and Interference Situation} \label{interferencecontextuality}
If we consider Table 3, we see that the experimental values measured by \cite{hampton1988b} for a whole number of items cannot be modeled within our quantum modeling scheme making use of the effect of interference. In \cite{aertsgabora2005a} and \cite{aertsgabora2005b} we have studied how the effect of context can be modeled within our quantum modeling scheme. We have also made ample measurements of the context effects on the state of  concepts, and shown how context many times exists of other concepts. From the type of experiments that we have performed to collect our data in \cite{aertsgabora2005a} it is plausible to suppose that also in the experiments considered in \cite{hampton1988b} the effect of context has played a role. Hence, what we mean concretely is the following. Suppose that for the pair of concepts {\it Fruits} and {\it Vegetables} and their disjunction {\it Fruits or Vegetables}, described in our quantum modeling scheme by states $|A\rangle$, $|B\rangle$, and $|AB\rangle$ of the Hilbert space ${\cal H}$, respectively, the membership weights $\mu(A)$, $\mu(B)$ and $\mu(A\ {\rm or}\ B)$ for an item {\it Olive} are considered. Then, before the subject takes the decision in favor or against membership, there is already an influence of context taking place due to their considering the specific item {\it Olive}. And this context effect will eventually be different for each of the considered items $X$, meaning that if for the same pair of concepts {\it Fruits} and {\it Vegetables} another item, for example {\it Apple}, is considered with respect to membership or non membership, the effect of considering the item {\it Apple} will introduce another context effect than in the case of considering {\it Olive}.

\subsection{The Joint Context and Interference Effect}

Let us describe such a joint effect of context and interference within our quantum modeling scheme. Hence, also the context effect will be modeled by an orthogonal projection $N$, that projects the states $|A\rangle$ and $|B\rangle$ describing the concepts $A$ and $B$ into new vectors $N|A\rangle$ and $N|B\rangle$. Also the superposition vector ${1 \over \sqrt{2}}(|A\rangle+|B\rangle)$, describing the concept `$A$ or $B$' will be projected and because of the linearity of the projection operation it will remain a superposition of the projected vectors of $|A\rangle$ and $|B\rangle$. More specifically we have
\begin{equation}
N{1 \over \sqrt{2}}(|A\rangle+|B\rangle)={1 \over \sqrt{2}}(N|A\rangle+N|B\rangle)
\end{equation}
The new states of the concepts $A$, $B$ and `$A$ or $B$' after this effect of context related to the item $X$, let us denote them $|A^N\rangle$, $|B^N\rangle$ and $|AB^N\rangle$, are described by the vectors encountered by normalizing these projected vectors, hence
\begin{equation} \label{Nvectors}
|A^N\rangle={N|A\rangle \over |N|A\rangle|},  \quad |B^N\rangle={N|B\rangle \over |N|B\rangle|} \quad |AB^N\rangle={{1 \over \sqrt{2}}(N|A\rangle+N|B\rangle) \over |{1 \over \sqrt{2}}(N|A\rangle+N|B\rangle)|}
\end{equation}
After the effect of context, described by the action of the orthogonal projection $N$, we can now proceed and describe the effect of interference due to the decision measurement in favor or against membership of the item $X$ with respect to the concepts $A$, $B$ and `$A$ or $B$'. This decision measurement is described by the orthogonal projection $M$, and hence the membership weights $\mu(A)$, $\mu(B)$ and $\mu(A\ {\rm or}\ B)$ are now given by
\begin{eqnarray}
\mu(A)&=&{\langle A|N|M|N|A\rangle \over \langle A|N|A\rangle} \quad \mu(B)={\langle B|N|M|N|B\rangle \over \langle B|N|B\rangle} \\
\mu(A\ {\rm or}\ B)&=&{({1 \over 2}\langle A|+\langle B|)N|M|N(|A\rangle+|B\rangle) \over ({1 \over 2}\langle A|+\langle B|)N(|A\rangle+|B\rangle)}
\end{eqnarray} 
Making use of $(|N|A\rangle|)^2=\langle A|N|A\rangle$, $(|N|B\rangle|)^2=\langle B|N|B\rangle$ and $(|{1 \over \sqrt{2}}(N|A\rangle+N|B\rangle)|)^2=({1 \over 2}\langle A|+\langle B|)N(|A\rangle+|B\rangle)$, we find
\begin{equation}
\mu(A\ {\rm or}\ B)={\langle A|N|M|N|A\rangle+\langle B|N|M|N|B\rangle+2\Re(\langle A|N|M|N|B\rangle) \over \langle A|N|A\rangle+\langle B|N|B\rangle+2\Re(\langle A|N|B\rangle)}
\end{equation}
We suppose that the context effect consisting of considering the item $X$ and the decision measurement consisting of choosing in favor or against membership of item $X$ are compatible. In quantum mechanics compatibility is expressed mathematically by the corresponding projection operators commuting, hence $MN=NM$. It can be shown -- this is just a verification of (\ref{linearity}), (\ref{hermiticity}) and (\ref{idempotenty}) -- that in this case also $MN$ is a projector, and hence we have
\begin{eqnarray} \label{mu(AorB)original}
\mu(A\ {\rm or}\ B)&=&{\langle A|MN|A\rangle+\langle B|MN|B\rangle+2\Re(\langle A|MN|B\rangle) \over \langle A|N|A\rangle+\langle B|N|B\rangle+2\Re(\langle A|N|B\rangle)}
\end{eqnarray}
Consider the following three orthogonal subspaces of the Hilbert space $MN({\cal H})$, $(1-M)N({\cal H})$ and $(1-N)({\cal H})$ we have
\begin{equation}
{\cal H}=N({\cal H})\oplus(1-N)({\cal H})=MN({\cal H})\oplus(1-M)N({\cal H})\oplus(1-N)({\cal H})
\end{equation}
and hence we can write
\begin{eqnarray}
|A\rangle&=&ae^{i\alpha}|e\rangle+a'e^{i\alpha'}|e'\rangle+\tilde{a}e^{i\tilde{\alpha}}|\tilde{e}\rangle \\
|B\rangle&=&be^{i\beta}|f\rangle+b'e^{i\beta'}|f'\rangle+\tilde{b}e^{i\tilde{\beta}}|\tilde{f}\rangle
\end{eqnarray}
such that $\langle e|e\rangle=\langle e'|e'\rangle=\langle \tilde{e}|\tilde{e}\rangle=1$ and $\langle e|e'\rangle=\langle e|\tilde{e}\rangle=\langle e'|\tilde{e}\rangle=0$. If we put $\langle e|f\rangle=ce^{i\gamma}$, $\langle e'|f'\rangle=c'e^{i\gamma'}$ and $\langle \tilde{e}|\tilde{f}\rangle=\tilde{c}e^{i\tilde{\gamma}}$ we get
\begin{eqnarray}
&\langle A|MN|A\rangle=a^2 \quad \langle B|MN|B\rangle=b^2 \quad \langle A|MN|B\rangle=abce^{i(\beta-\alpha+\gamma)} \\
&\langle A|N|A\rangle=a^2+a'^2 \quad \langle B|N|B\rangle=b^2+b'^2 \\
&\langle A|N|B\rangle=abce^{i(\beta-\alpha+\gamma)}+a'b'c'e^{i(\beta'-\alpha'+\gamma')}
\end{eqnarray}
Putting $\phi=\beta-\alpha+\gamma$, $\phi'=\beta'-\alpha'+\gamma'$ and $\tilde{\phi}=\tilde{\beta}-\tilde{\alpha}+\tilde{\gamma}$ we have
\begin{eqnarray}
&&\mu(A)={a^2 \over a^2+a'^2} \quad \mu(B)={b^2 \over b^2+b'^2} \\
&&\mu(A\ {\rm or}\ B)={a^2+b^2+2abc\cos\phi \over a^2+a'^2+b^2+b'^2+2abc\cos\phi+2a'b'c'\cos\phi'}
\end{eqnarray}
Let us introduce 
\begin{equation}
n^2=a^2+a'^2 \quad n'^2=b^2+b'^2
\end{equation}
then we have 
\begin{eqnarray} \label{aba'b'tildeatildeb}
&a^2=n^2\mu(A) \quad b^2=n'^2\mu(B) \\
&a'^2=n^2(1-\mu(A)) \quad b'^2=n'^2(1-\mu(B)) \\ &\tilde{a}^2=1-n^2 \quad \tilde{b}^2=1-n'^2
\end{eqnarray}
and hence
\begin{eqnarray} \label{mu(AorB)}
&&\mu(A\ {\rm or}\ B)= \\
&&{n^2\mu(A)+n'^2\mu(B)+2nn'\sqrt{\mu(A)\mu(B)}c\cos\phi \over n^2+n'^2+2nn'\sqrt{\mu(A)\mu(B)}c\cos\phi+2nn'\sqrt{(1-\mu(A))(1-\mu(B)}c'\cos\phi'} \nonumber
\end{eqnarray}
Since $\langle A|B\rangle=0$ we have
\begin{eqnarray}
0&=&abe^{i(\beta-\alpha)}\langle e|f\rangle+a'b'e^{i(\beta'-\alpha')}\langle e'|f'\rangle+\tilde{a}\tilde{b}e^{i(\tilde{\beta}-\tilde{\alpha})}\langle \tilde{e}|\tilde{f}\rangle \\
&=&abce^{i(\beta-\alpha+\gamma)}+a'b'c'e^{i(\beta'-\alpha'+\gamma')}+\tilde{a}\tilde{b}\tilde{c}e^{i(\tilde{\beta}-\tilde{\alpha}+\tilde{\gamma})} \\ \label{ABorthogonal}
&=&abce^{i\phi}+a'b'c'e^{i\phi'}+\tilde{a}\tilde{b}\tilde{c}e^{i\tilde{\phi}}
\end{eqnarray}
We prove now that for an arbitrary $\mu(A)$, $\mu(B)$ and $\mu(A\ {\rm or}\ B)$, we can find a quantum representation that models these data, combining the effects of context and interference within the modeling scheme we put forward. Our proof will consist in constructing an explicit solution for such an arbitrary set of data, and more specifically, we will show that we can construct a solution for which $c=c'=\tilde{c}=1$ and $\phi'=\phi+\pi$. The equations that we need to satisfy to find a solution are equations (\ref{mu(AorB)}) and (\ref{ABorthogonal}). After substituting the values of $c$, $c'$, $\tilde{c}$ and $\phi'$, and also making use of (\ref{aba'b'tildeatildeb}), they reduce to the following two equations
\begin{eqnarray} \label{sol01}
&&\mu(A\ {\rm or}\ B)= \\
&&{n^2\mu(A)+n'^2\mu(B)+2nn'\sqrt{\mu(A)\mu(B)}\cos\phi \over n^2+n'^2+2nn'\sqrt{\mu(A)\mu(B)}\cos\phi-2nn'\sqrt{(1-\mu(A))(1-\mu(B)}\cos\phi} \nonumber \\
\label{sol02}
&&\sqrt{(1-n^2)(1-n'^2)}e^{i\tilde{\theta}}=nn'R
\end{eqnarray}
with
\begin{equation}
R=\sqrt{(1-\mu(A))(1-\mu(B))}-\sqrt{\mu(A)\mu(B)}
\end{equation}
Let us first explain how we choose to satisfy equation (\ref{sol02}). For an arbitrary $\mu(A)$ and $\mu(B)$, we can have (i) $\mu(A)+\mu(B)\le1$ or (ii) $1<\mu(A)+\mu(B)$. In case (i), we have $\mu(A)\mu(B)\le (1-\mu(A))(1-\mu(B))$ and hence $\sqrt{\mu(A)\mu(B)}\le \sqrt{(1-\mu(A))(1-\mu(B))}$ or $0\le R$. In case (ii), we have $R<0$. We will choose the phase $\tilde{\theta}$ according to these two cases (i) and (ii). More specifically, we choose $\tilde{\theta}=0$ in case (i) and $\tilde{\theta}=\pi$ in case (ii). With this choice of $\tilde{\theta}$, equation (\ref{sol02}) becomes
\begin{equation} \label{sol02bis}
\sqrt{(1-n^2)(1-n'^2)}=nn'|R|
\end{equation}
where $|R|$ is the absolute value of $R$. We introduce the following three intervals 
\begin{eqnarray}
I_u&=&[0,\min(\mu(A),\mu(B))] \\
I_c&=&]\min(\mu(A),\mu(B)),\max(\mu(A),\mu(B))[ \\ I_o&=&[\max(\mu(A),\mu(B)),1]
\end{eqnarray}
constituting a partition of the interval $[0,1]$, and call them  the `underextension interval', the `context interval', and the `overextension interval', respectively, referring to the situation for $\mu(A\ {\rm or}\ B)$ pertaining to each one of these intervals.

\subsection{The situation of convexity}

Let us consider first the situation where $\mu(A\ {\rm or}\ B) \in I_c=]\min(\mu(A),\mu(B)),$ $\max(\mu(A),\mu(B))[$. In this case $\mu(A\ {\rm or}\ B)$ is a convex combination of $\mu(A)$ and $\mu(B)$. This value of $\mu(A\ {\rm or}\ B)$ can be modeled in our theory by choosing $\phi={\pi \over 2}$. For this choice of $\phi$ we have $\cos\phi=0$ and hence (\ref{sol01}) becomes
\begin{equation} \label{mu(AorB)convex}
\mu(A\ {\rm or}\ B)={n^2\mu(A)+n'^2\mu(B) \over n^2+n'^2}={n^2 \over n^2+n'^2}\mu(A)+{n'^2 \over n^2+n'^2}\mu(B)
\end{equation}
which shows that $\mu(A\ {\rm or}\ B)$ is a convex combination of $\mu(A)$ and $\mu(B)$, and hence $\mu(A\ {\rm or}\ B)\in I_c$. Moreover for an arbitrary $\mu(A\ {\rm or}\ B)$, we can calculate $n'$ in function of $n$ using (\ref{mu(AorB)convex}). This gives 
\begin{eqnarray}
&&(n^2+n'^2)\mu(A\ {\rm or}\ B)=n^2\mu(A)+n'^2\mu(B) \\
&\Leftrightarrow&n'^2(\mu(A\ {\rm or}\ B)-\mu(B))=n^2(\mu(A\ {\rm or}\ B)-\mu(A) \\ 
&\Leftrightarrow&n'^2=n^2({\mu(A\ {\rm or}\ B)-\mu(A) \over \mu(B)-\mu(A\ {\rm or}\ B)})
\end{eqnarray}
in case $\mu(A\ {\rm or}\ B)\not=\mu(B)$, which is always satisfied since $\mu(A\ {\rm or}\ B)\in I_c$. Substituting the value of $n'$ in (\ref{sol02bis}) we get
\begin{eqnarray}
&&(1-n^2)(1-n'^2)=n^2n'^2R^2 \\
&\Leftrightarrow&1-n^2-n'^2+n^2n'^2(1-R^2)=0 \\
&\Leftrightarrow&1-n^2(1+{\mu(A\ {\rm or}\ B)-\mu(A) \over \mu(B)-\mu(A\ {\rm or}\ B)}) \nonumber \\
&&+n^4({\mu(A\ {\rm or}\ B)-\mu(A) \over \mu(B)-\mu(A\ {\rm or}\ B)})(1-R^2)=0 \\
&\Leftrightarrow&
f_c(n)=0
\end{eqnarray}
with
\begin{equation}
f_c(n)=\mu(B)-\mu(A\ {\rm or}\ B)-n^2(\mu(B)-\mu(A))+n^4(\mu(A\ {\rm or}\ B)-\mu(A))(1-R^2)
\end{equation}
For the function $f_c$ we have $f_c(0)=$ \ $\mu(B)-\mu(A\ {\rm or}\ B)$ and $f_c(1)=-R^2(\mu(A\ {\rm or}\ B)-\mu(A))$, which means that $f_c(0)$ and $f_c(1)$ have a different sign. Since $f_c$ is a continuous function in the variable $n$ it becomes $0$ for a value of $n$ in the interval $[0,1]$. This value is the solution we look for. Let us determine it explicitly. In case $\mu(A\ {\rm or}\ B)\not=\mu(A)$, which is always satisfied since $\mu(A\ {\rm or}\ B)\in I_c$, the function $f_c(n)$ is a quadratic function in $n^2$ of which the discriminant is given by
\begin{equation}
D_c=(\mu(B)-\mu(A))^2-4(\mu(A\ {\rm or}\ B)-\mu(A))(\mu(B)-\mu(A\ {\rm or}\ B))(1-R^2)
\end{equation}
and hence we get for $n$ and $n'$ the following solution
\begin{eqnarray}
n&=&\sqrt{\mu(B)-\mu(A)\pm\sqrt{D_c} \over 2(\mu(A\ {\rm or}\ B)-\mu(A))(1-R^2)} \\
n'&=&\sqrt{\mu(B)-\mu(A)\pm\sqrt{D_c} \over 2(\mu(A\ {\rm or}\ B)-\mu(B))(1-R^2)}
\end{eqnarray}
Before we proceed to consider the other situations we put forward the following lemma.

\bigskip
\noindent
{\bf Lemma:} Suppose we have $0\le u \le v$, $0\le w_1\le w_2$ and $0\le t_1\le t_2$ {\it (i) In case $0\le u+w_1 \le v+w_1-t_1$ and $0\le u+w_2 \le v+w_2-t_2$ we have
\begin{equation} \label{lemma01}
0 \le {u+w_1 \over v+w_1-t_1}\le{u+w_2 \over v+w_2-t_2} \le 1
\end{equation}
(ii) In case $0\le u-w_1\le v-w_1+t_1$ and $0\le u-w_2\le v-w_2+t_2$ we have}
\begin{equation} \label{lemma02}
0\le {u-w_2 \over v-w_2+t_2}\le{u-w_1 \over v-w_1+t_1}\le1
\end{equation}
Proof: Suppose that $u, v, w_1, w_2, t_1, t_2$ fulfill the conditions of the lemma and of (i). From $0\le u+w_1 \le v+w_1-t_1$ and $0\le u+w_2 \le v+w_2-t_2 \le v+w_2-t_1$ follows that
\begin{equation}
0\le{u+w_1 \over v+w_1-t_1} \le 1 \quad {\rm and} \quad 0\le{u+w_2 \over v+w_2-t_1} \le 1
\end{equation}
From this follows that
\begin{eqnarray}
&&0\le1-{u+w_2 \over v+w_2-t_1}={v+w_2-t_1-u-w_2 \over v+w_2-t_1}={v-t_1-u \over v+w_2-t_1} \nonumber \\
&&\le{v-t_1-u \over v+w_1-t_1}={v+w_1-t_1-u-w_1 \over v+w_1-t_1}=1-{u+w_1 \over v+w_1-t_1}\le1
\end{eqnarray}
and hence
\begin{equation}
0\le{u+w_1 \over v+w_1-t_1}\le{u+w_2 \over v+w_2-t_1}\le{u+w_2 \over v+w_2-t_2}\le1
\end{equation}
which proves (\ref{lemma01}). Suppose now that $u, v, w_1, w_2, t_1, t_2$ fulfill the conditions of the lemma and of (ii). From $0\le u-w_1\le v-w_1+t_1\le v-w_1+t_2$ and $0\le u-w_2\le v-w_2+t_2$ follows that
\begin{equation}
0\le {u-w_2 \over v-w_2+t_2}\le1 \quad {\rm and} \quad 0\le{u-w_1 \over v-w_1+t_2}\le1
\end{equation}
From this follows that
\begin{eqnarray}
&&0\le 1-{u-w_1 \over v-w_1+t_2}={v-w_1+t_2-u+w_1 \over v-w_1+t_2}={v+t_2-u \over v-w_1+t_2} \nonumber \\
&&\le{v+t_2-u \over v-w_2+t_2}={v-w_2+t_2-u+w_2 \over v-w_2+t_2}=1-{u-w_2 \over v-w_2+t_2}\le1
\end{eqnarray}
and hence
\begin{equation}
0\le{u-w_2 \over v-w_2+t_2}\le{u-w_1 \over v-w_1+t_2}\le{u-w_1 \over v-w_1+t_1}\le1
\end{equation}
which proves (\ref{lemma02}).

\subsection{The situation of double overextension}

Let us consider the situation where $\mu(A)\le\mu(A\ {\rm or}\ B)$ and $\mu(B)\le\mu(A\ {\rm or}\ B)$. Take $\phi_1, \phi_2\in [0,{\pi \over 2}]$ such that $\phi_2\le\phi_1$, then $\cos\phi_1 \le \cos\phi_2$ and if we put $w_1=2nn'\sqrt{\mu(A)\mu(B)}\cos\phi_1$ and $w_2=2nn'\sqrt{\mu(A)\mu(B)}\cos\phi_1$ we have $0\le w_1 \le w_2$. Further put $u=n^2\mu(A)+n'^2\mu(B)$ and $v=n^2+n'^2$, then we have $0\le u \le v$. Put $t_1=2nn'\sqrt{(1-\mu(A))(1-\mu(B)}\cos\phi_1$ and $t_2=2nn'\sqrt{(1-\mu(A))(1-\mu(B)}\cos\phi_2$ and this gives $0\le t_1 \le t_2$. With these choices of $u$, $v$, $w_1$, $w_2$, $t_1$ and $t_2$ we also have $0\le u+w_1 \le v+w_1-t_1$ and $0\le u+w_2 \le v+w_2-t_2$, which means that all requirements of part (i) of the lemma are satisfied. As a consequence from this lemma follows that
\begin{eqnarray}
&&0\le \nonumber \\
&&{n^2\mu(A)+n'^2\mu(B)+2nn'\sqrt{\mu(A)\mu(B)}\cos\phi_1 \over n^2+n'^2+2nn'\sqrt{\mu(A)\mu(B)}\cos\phi_1-2nn'\sqrt{(1-\mu(A))(1-\mu(B)}\cos\phi_1}\le \nonumber \\
&&{n^2\mu(A)+n'^2\mu(B)+2nn'\sqrt{\mu(A)\mu(B)}\cos\phi_2 \over n^2+n'^2+2nn'\sqrt{\mu(A)\mu(B)}\cos\phi_2-2nn'\sqrt{(1-\mu(A))(1-\mu(B)}\cos\phi_2} \nonumber \\
&&\le1
\end{eqnarray}
This means that in function of $\phi$ the maximum of $\mu(A\ {\rm or}\ B)$ is reached for $\phi=0$, and because $\mu(A\ {\rm or}\ B)$ as expressed by equation (\ref{sol01}) is a continuous function of $\phi$, for decreasing values of $\phi$ we have that $\mu(A\ {\rm or}\ B)$ increases from ${n^2\mu(A)+n'^2\mu(B) \over n^2+n'^2}$, which is a value contained in $[\min(\mu(A),\mu(B)),$ $\max(\mu(A),\mu(B))]$, to this maximum. If we can prove that it is possible to choose $n$ and $n'$ such that this maximum equals 1, then all values of the interval $[{n^2\mu(A)+n'^2\mu(B) \over n^2+n'^2},1]$ are reached for intermediate values of $\phi$. This means that a value of $\phi$ exists that models any double overextended value of $\mu(A\ {\rm or}\ B)$. Let us prove this. The maximum is reached for $\phi=0$ and to determine $n$ and $n'$ such that this maximum equals 1, we need to have
\begin{eqnarray} \label{mu(AorB)=1}
&&\!\!\!\!\!\!\!\! 1={n^2\mu(A)+n'^2\mu(B)+2nn'\sqrt{\mu(A)\mu(B)} \over n^2+n'^2+2nn'\sqrt{\mu(A)\mu(B)}-2nn'\sqrt{(1-\mu(A))(1-\mu(B)}} \\
&\Leftrightarrow&n^2+n'^2+2nn'\sqrt{\mu(A)\mu(B)}-2nn'\sqrt{(1-\mu(A))(1-\mu(B)} \nonumber \\
&&=n^2\mu(A)+n'^2\mu(B)+2nn'\sqrt{\mu(A)\mu(B)} \\
&\Leftrightarrow&n^2+n'^2-2nn'\sqrt{(1-\mu(A))(1-\mu(B)}=n^2\mu(A)+n'^2\mu(B) \\
&\Leftrightarrow&n^2(1-\mu(A))+n'^2(1-\mu(B)) \nonumber \\
&&-2nn'\sqrt{(1-\mu(A))(1-\mu(B)}=0 \\
&\Leftrightarrow&(n\sqrt{1-\mu(A)}-n'\sqrt{1-\mu(B)})^2=0 \\ \label{nn'relation}
&\Leftrightarrow&n\sqrt{1-\mu(A)}=n'\sqrt{1-\mu(B)} \\
&\Leftrightarrow&n'^2=n^2{1-\mu(A) \over 1-\mu(B)}
\end{eqnarray}
in case $\mu(B)\not=1$. Substituting the value of $n'$ in (\ref{sol02bis}) we get
\begin{eqnarray}
&&(1-n^2)(1-n'^2)=n^2n'^2R^2 \\
&\Leftrightarrow&1-n^2(1+{1-\mu(A) \over 1-\mu(B)})+n^4{1-\mu(A) \over 1-\mu(B)}(1-R^2)=0 \\ \label{sol06}
&\Leftrightarrow&f_o(n)=0
\end{eqnarray}
with
\begin{equation}
f_o(n)=1-\mu(B)-n^2(2-\mu(A)-\mu(B))+n^4(1-\mu(A))(1-R^2)
\end{equation}
We have $f_o(0)=1-\mu(B)$ and hence $0\le f_o(0)$ and $f_o(1)=-(1-\mu(A)+(1-\mu(A))(1-R^2)=-(1-\mu(A))R^2$ and hence $f_o(1)\le0$. Since $f_o$ is a continuous function of $n$, it needs to become $0$ in the interval $[0,1]$. If $\mu(A)\not=1$ the function $f_o(n)$ is a quadratic function in $n^2$ of which the discriminant is given by
\begin{equation}
D_o=(2-\mu(A)-\mu(B))^2-4(1-\mu(A))(1-\mu(B))(1-R^2)
\end{equation}
and hence we get for $n$ and $n'$ the following solution
\begin{eqnarray}
n&=&\sqrt{{2-\mu(A)-\mu(B)\pm\sqrt{D_o} \over 2(1-\mu(A))(1-R^2)}} \\
n'&=&\sqrt{{2-\mu(A)-\mu(B)\pm\sqrt{D_o} \over 2(1-\mu(B))(1-R^2)}}
\end{eqnarray}
In case $\mu(B)=1$ from (\ref{nn'relation}) follows then that $n=0$ and from (\ref{sol02bis}) follows then $n'=1$. In case $\mu(A)=1$ we get from (\ref{nn'relation}) and (\ref{sol02bis}) that $n=1$ and $n'=0$. In case $\mu(B)=\mu(A)=1$ we have that (\ref{nn'relation}) is satisfied for all values of $n$ and $n'$ such that only (\ref{sol02bis}) needs to be satisfied, which leaves different values of $n$ and $n'$ possible. The case where $\mu(A)=\mu(B)=0$ we will consider apart in detail in section \ref{compl3representation}.

The values of $n$, $n'$ that we calculated correspond to the situation where $\mu(A\ {\rm or}\ B)=1$. Then we determine $\theta$ such that we find the experimental value for $\mu(A\ {\rm or}\ B)$ in the following way. We need to have
\begin{eqnarray}
&&\!\!\!\!\!\!\!\! \mu(A\ {\rm or}\ B)= \\
&&\!\!\!\!\!\!\!\! {n^2\mu(A)+n'^2\mu(B)+2nn'\sqrt{\mu(A)\mu(B)}\cos\phi \over n^2+n'^2+2nn'\sqrt{\mu(A)\mu(B)}\cos\phi-2nn'\sqrt{(1-\mu(A))(1-\mu(B)}\cos\phi} \nonumber \\
&\Leftrightarrow& \mu(A\ {\rm or}\ B)(n^2+n'^2+2nn'\sqrt{\mu(A)\mu(B)}\cos\phi \nonumber \\
&&-2nn'\sqrt{(1-\mu(A))(1-\mu(B)}\cos\phi) \nonumber \\
&&=n^2\mu(A)+n'^2\mu(B)+2nn'\sqrt{\mu(A)\mu(B)}\cos\phi \\
&\Leftrightarrow& n^2(\mu(A\ {\rm or}\ B)-\mu(A))+n'^2(\mu(A\ {\rm or}\ B)-\mu(B)) \nonumber \\
&&=(2nn'(1-\mu(A\ {\rm or}\ B))\sqrt{\mu(A)\mu(B)} \nonumber \\
&&+2nn'\mu(A\ {\rm or}\ B)\sqrt{(1-\mu(A))(1-\mu(B)})\cos\phi \\
&\Leftrightarrow& \cos\phi= \\
&&\!\!\!\!\!\!\!\! {n^2(\mu(A\ {\rm or}\ B)-\mu(A))+n'^2(\mu(A\ {\rm or}\ B)-\mu(B)) \over 2nn'(1-\mu(A\ {\rm or}\ B))\sqrt{\mu(A)\mu(B)}+2nn'\mu(A\ {\rm or}\ B)\sqrt{(1-\mu(A))(1-\mu(B)}} \nonumber
\end{eqnarray}
This results in
\begin{equation} \label{phioverextension}
{\textstyle \phi=\arccos({n^2(\mu(A\ {\rm or}\ B)-\mu(A))+n'^2(\mu(A\ {\rm or}\ B)-\mu(B)) \over 2nn'(1-\mu(A\ {\rm or}\ B))\sqrt{\mu(A)\mu(B)}+2nn'\mu(A\ {\rm or}\ B)\sqrt{(1-\mu(A))(1-\mu(B)}})}
\end{equation}
This is well defined, since only if of $\mu(A)$ and $\mu(B)$ one equals 1 and the other equals 0 there could be a problem. But in this case we have $\mu(A\ {\rm or}\ B)=1$, and hence the situation was already treated, with as result that of $n$ and $n'$ one equals 0 and the other equals 1, and $\phi$ can be chosen arbitrarily since both (\ref{nn'relation}) and (\ref{sol02bis}) are satisfied.

\subsection{The situation of double underextension}

Let us consider next the situation where $\mu(A\ {\rm or}\ B)\le\mu(A)$ and $\mu(A\ {\rm or}\ B)\le\mu(B)$. Take $\phi_1, \phi_2\in [{\pi \over 2},\pi]$ such that $\phi_1\le\phi_2$, then $-\cos\phi_1 \le -\cos\phi_2$ and if we put $w_1=-2nn'\sqrt{\mu(A)\mu(B)}\cos\phi_1$ and $w_2=-2nn'\sqrt{\mu(A)\mu(B)}\cos\phi_1$ we have $0\le w_1 \le w_2$. Further put $u=n^2\mu(A)+n'^2\mu(B)$ and $v=n^2+n'^2$, then we have $0\le u \le v$. Put $t_1=-2nn'\sqrt{(1-\mu(A))(1-\mu(B)}\cos\phi_1$ and $t_2=-2nn'\sqrt{(1-\mu(A))(1-\mu(B)}\cos\phi_2$ and this gives $0\le t_1 \le t_2$. With these choices of $u$, $v$, $w_1$, $w_2$, $t_1$ and $t_2$ we also have $0\le u-w_1 \le v-w_1+t_1$ and $0\le u-w_2 \le v-w_2+t_2$, which means that all requirements of part (ii) of the lemma are satisfied. As a consequence from this lemma follows that
\begin{eqnarray}
&&0\le \nonumber \\
&&{n^2\mu(A)+n'^2\mu(B)+2nn'\sqrt{\mu(A)\mu(B)}\cos\phi_2 \over n^2+n'^2+2nn'\sqrt{\mu(A)\mu(B)}\cos\phi_2-2nn'\sqrt{(1-\mu(A))(1-\mu(B)}\cos\phi_2} \nonumber \\
&&\le{n^2\mu(A)+n'^2\mu(B)+2nn'\sqrt{\mu(A)\mu(B)}\cos\phi_1 \over n^2+n'^2+2nn'\sqrt{\mu(A)\mu(B)}\cos\phi_1-2nn'\sqrt{(1-\mu(A))(1-\mu(B)}\cos\phi_1} \nonumber \\
&&\le1
\end{eqnarray}
This means that in function of $\phi$ the minimum of $\mu(A\ {\rm or}\ B)$ is reached for $\phi=\pi$, and because $\mu(A\ {\rm or}\ B)$ as expressed by equation (\ref{sol01}) is a continuous function of $\phi$, for increasing values of $\phi$ we have that $\mu(A\ {\rm or}\ B)$ decreases from ${n^2\mu(A)+n'^2\mu(B) \over n^2+n'^2}$, which is a value contained in $[\min(\mu(A),\mu(B)),$ $\max(\mu(A),\mu(B))]$, to this minimum. If we can prove that it is possible to choose $n$ and $n'$ such that this minimum equals 0, then all values of the interval $[0,{n^2\mu(A)+n'^2\mu(B) \over n^2+n'^2}]$ are reached for intermediate values of $\phi$. This means that a value of $\phi$ exists that models any double underextended value of $\mu(A\ {\rm or}\ B)$. Let us prove this. The minimum is reached for $\phi=0$ and to determine $n$ and $n'$ such that this minimum equals 0, we need to have
\begin{eqnarray}
&&0=n^2\mu(A)+n'^2\mu(B)-2nn'\sqrt{\mu(A)\mu(B)} \\
&\Leftrightarrow&(n\sqrt{\mu(A)}-n'\sqrt{\mu(B)})^2=0 \\ \label{nn'relationbis}
&\Leftrightarrow&n\sqrt{\mu(A)}=n'\sqrt{\mu(B)} \\
&\Leftrightarrow&n'^2=n^2{\mu(A) \over \mu(B)}
\end{eqnarray}
in case $\mu(B)\not=0$. Further we need (\ref{sol02bis}) to be satisfied, and hence
\begin{eqnarray} \label{sol05bis}
&&\sqrt{(1-n^2)(1-n'^2)}=nn'R \\
&\Leftrightarrow&(1-n^2)(1-n'^2)=n^2n'^2R^2 \\
&\Leftrightarrow&1-n^2-n'^2+n^2n'^2(1-R^2)=0 \\
&\Leftrightarrow&1-n^2(1+{\mu(A) \over \mu(B)})+n^4{\mu(A) \over \mu(B)}(1-R^2)=0 \\ \label{sol06bis}
&\Leftrightarrow&g_u(n)=\mu(B)-n^2(\mu(A)+\mu(B))+n^4\mu(A)(1-R^2)=0
\end{eqnarray}
We have $g_u(0)=\mu(B)$ and hence $0\le g_u(0)$ and $g_u(1)=-\mu(A)+\mu(A)(1-R^2)=-\mu(A)R^2$ and hence $g_u(1)\le0$. Since $g_u$ is a continuous function of $n$, it needs to become $0$ in the interval $[0,1]$. If $\mu(A)\not=0$, (\ref{sol06bis}) is a quadratic equation, and hence we can calculate this solution explicitly. The discriminant of the equation is given by
\begin{equation}
D_u=(\mu(A)+\mu(B))^2-4\mu(A)\mu(B)(1-R^2) \\
\end{equation}
and values of $n$ and $n'$ are given by
\begin{eqnarray}
n&=&\sqrt{{\mu(A)+\mu(B)\pm\sqrt{D_u} \over 2\mu(A)(1-R^2)}} \\
n'&=&\sqrt{{\mu(A)+\mu(B)\pm\sqrt{D_u} \over 2\mu(B)(1-R^2)}}
\end{eqnarray}
In case $\mu(A)=0$ from (\ref{nn'relationbis}) follows that $n'=0$ and from (\ref{sol02bis}) follows then $n=1$. In case $\mu(B)=0$ from (\ref{nn'relationbis}) and (\ref{sol02bis}) follows then that $n=0$ and $n'=1$. In case $\mu(A)=\mu(B)=0$ we have that (\ref{nn'relationbis}) is satisfied for all values of $n$ and $n'$ such that only (\ref{sol02bis}) needs to be satisfied, which leaves different values of $n$ and $n'$ possible.

The values of $n$, $n'$ that we calculated correspond to the situation where $\mu(A\ {\rm or}\ B)=0$. Let us determine $\theta$ such that we find the experimental value for $\mu(A\ {\rm or}\ B)$. For this we make an analogous calculation than the one for the situation of double overextension, and find 
\begin{equation}
{\textstyle \phi=\arccos({n^2(\mu(A\ {\rm or}\ B)-\mu(A))+n'^2(\mu(A\ {\rm or}\ B)-\mu(B)) \over 2nn'(1-\mu(A\ {\rm or}\ B))\sqrt{\mu(A)\mu(B)}+2nn'\mu(A\ {\rm or}\ B)\sqrt{(1-\mu(A))(1-\mu(B)}})}
\end{equation}
This is well defined, since only if of $\mu(A)$ and $\mu(B)$ one equals 1 and the other equals 0 there could be a problem. But in this case we have $\mu(A\ {\rm or}\ B)=0$, and hence the situation was already treated, with as result that of $n$ and $n'$ one equals 0 and the other equals 1, and $\phi$ can be chosen arbitrarily, since both (\ref{nn'relationbis}) and (\ref{sol02bis}) are satisfied.

\section{Construction of a $\compl^3$ Representation} \label{compl3representation}
In this section we construct an explicit representation in a three dimensional complex Hilbert space for the experimental data of \cite{hampton1988b}, taking into account our analysis of the foregoing section. Hence, this representation combines an effect of context, quantum mechanically modeled by means of the orthogonal projection $N$, with an effect of interference due to a decision, quantum mechanically modeled by means of the orthogonal projection $M$.

\subsection{Double Onderextension, Convexity and Double Overextension}

Hence, suppose we consider two concepts $A$ and $B$ and their disjunction `$A$ or $B$', and $\mu(A)$, $\mu(B)$ and $\mu(A\ {\rm or}\ B)$ are the measured membership weights for an item $X$ with respect to $A$, $B$ and `$A$ or $B$'. We model these weights within our quantum modeling scheme as follows. Consider the three dimensional Hilbert space $\compl^3$. Concepts $A$, $B$ and `$A$ or $B$' are represented by vectors $|A\rangle$, $|B\rangle$ and ${1 \over \sqrt{2}}(|A\rangle+|B\rangle)$ of $\compl^3$. The context effect of considering the item $X$ with respect to $A$, $B$ and `$A$ or $B$' for the subject being tested is described by the orthogonal projection $N$ that maps all vectors orthogonally onto the plane generated by $(1,0,0)$ and $(0,1,0)$. The decision of a subject to decide in favor of membership for the item $X$ with respect to concepts $A$, $B$ and `$A$ or $B$' is described by the orthogonal projection $M$ that maps all vectors orthogonally onto the ray generated by the vector $(1,0,0)$, and the decision of the subject against membership for the item $X$ with respect to concepts $A$, $B$ and `$A$ or $B$' is described by the orthogonal projection $M$ that maps all vectors orthogonally onto the ray generated by the vector $(0,1,0)$. To define more explicitly the vectors $|A\rangle$, $|B\rangle$ and ${1 \over \sqrt{2}}(|A\rangle+|B\rangle)$, we consider the following situations.

\bigskip
\noindent
{\it (i) The situation of double underextension}: Hence $\mu(A\ {\rm or}\ B)\le\mu(A)$ and $\mu(A\ {\rm or}\ B)\le\mu(B)$. In this situation we take
\begin{equation}
|A\rangle=(n\sqrt{\mu(A)},n\sqrt{1-\mu(A)},\sqrt{1-n^2})
\end{equation}
\begin{equation}
|B\rangle=e^{i\beta}(n'\sqrt{\mu(B)},-n'\sqrt{1-\mu(B)},\pm\sqrt{1-n^2})
\end{equation}
\begin{equation}
{\textstyle \beta=\arccos({n^2(\mu(A\ {\rm or}\ B)-\mu(A))+n'^2(\mu(A\ {\rm or}\ B)-\mu(B)) \over 2nn'(1-\mu(A\ {\rm or}\ B))\sqrt{\mu(A)\mu(B)}+2nn'\mu(A\ {\rm or}\ B)\sqrt{(1-\mu(A))(1-\mu(B)}})}
\end{equation}
If $\mu(A)\not=0$ and $\mu(B)\not=0$ we take
\begin{eqnarray}
n&=&\sqrt{{\mu(A)+\mu(B)\pm\sqrt{D_u} \over 2\mu(A)(1-R^2)}} \\
n'&=&\sqrt{{\mu(A)+\mu(B)\pm\sqrt{D_u} \over 2\mu(B)(1-R^2)}}
\end{eqnarray}
If $\mu(A)=0$ we take $n=1$ and $n'=0$, and if $\mu(B)=0$ we take $n=0$ and $n'=1$. Further we take
\begin{eqnarray}
&D_u=(\mu(A)+\mu(B))^2-4\mu(A)\mu(B)(1-R^2) \\
&R=\sqrt{(1-\mu(A))(1-\mu(B))}-\sqrt{\mu(A)\mu(B)}
\end{eqnarray}
where we use the plus sign in front of $\sqrt{1-n^2}$ in the third coordinate of $|B\rangle$ in case $\mu(A)+\mu(B)\le1$, and the minus sign in case $1<\mu(A)+\mu(B)$.

\bigskip
\noindent
{\it (ii) The situation of convexity}: Hence $\min(\mu(A),\mu(B))<\mu(A\ {\rm or}\ B)$ and $\mu(A\ {\rm or}\ B)<\max(\mu(A),\mu(B))$. In this situation we take
\begin{equation}
|A\rangle=(n\sqrt{\mu(A)},n\sqrt{1-\mu(A)},\sqrt{1-n^2})
\end{equation}
\begin{equation}
|B\rangle=e^{i\beta}(n'\sqrt{\mu(B)},-n'\sqrt{1-\mu(B)},\pm\sqrt{1-n^2})
\end{equation}
We take $\beta={\pi \over 2}$ and if $\mu(A)\not=\mu(A\ {\rm or}\ B)$ and $\mu(B)\not=\mu(A\ {\rm or}\ B)$ we take
\begin{eqnarray}
n&=&\sqrt{{\mu(B)-\mu(A)\pm\sqrt{D_c} \over 2(\mu(A\ {\rm or}\ B)-\mu(A))(1-R^2)}} \\
n'&=&\sqrt{{\mu(B)-\mu(A)\pm\sqrt{D_c} \over 2(\mu(A\ {\rm or}\ B)-\mu(B))(1-R^2)}}
\end{eqnarray}
and further
\begin{eqnarray}
&D_c=(\mu(B)-\mu(A))^2 \nonumber \\
&-4(\mu(A\ {\rm or}\ B)-\mu(A))(\mu(B)-\mu(A\ {\rm or}\ B))(1-R^2)
 \\
&R=\sqrt{(1-\mu(A))(1-\mu(B))}-\sqrt{\mu(A)\mu(B)} 
\end{eqnarray}
where we use the plus sign in front of $\sqrt{1-n^2}$ in the third coordinate of $|B\rangle$ in case $\mu(A)+\mu(B)\le1$, and the minus sign in case $1<\mu(A)+\mu(B)$

\bigskip
\noindent
{\it (iii) The situation of double overextension}: Hence $\mu(A)<\mu(A\ {\rm or}\ B)$ and $\mu(B)<\mu(A\ {\rm or}\ B)$. In this situation we take
\begin{equation}
|A\rangle=(n\sqrt{\mu(A)},n\sqrt{1-\mu(A)},\sqrt{1-n^2})
\end{equation}
\begin{equation}
|B\rangle=e^{i\beta}(n'\sqrt{\mu(B)},-n'\sqrt{1-\mu(B)},\pm\sqrt{1-n^2})
\end{equation}
\begin{equation}
{\textstyle \beta=\arccos({n^2(\mu(A\ {\rm or}\ B)-\mu(A))+n'^2(\mu(A\ {\rm or}\ B)-\mu(B)) \over 2nn'(1-\mu(A\ {\rm or}\ B))\sqrt{\mu(A)\mu(B)}+2nn'\mu(A\ {\rm or}\ B)\sqrt{(1-\mu(A))(1-\mu(B)}})}
\end{equation}
and if $\mu(A)\not=1$ and $\mu(B)\not=1$ we take
\begin{eqnarray}
n&=&\sqrt{{2-\mu(A)-\mu(B)\pm\sqrt{D_o} \over 2(1-\mu(A))(1-R^2)}} \\
n'&=&\sqrt{{2-\mu(A)-\mu(B)\pm\sqrt{D_o} \over 2(1-\mu(B))(1-R^2)}}
\end{eqnarray}
and if $\mu(A)=1$ we take $n=1$ and $n'=0$, and if $\mu(B)=1$ we take $n=0$ and $n'=1$. We further take
\begin{eqnarray}
D_o&=&(2-\mu(A)-\mu(B))^2-4(1-\mu(A))(1-\mu(B))(1-R^2) \\
R&=&\sqrt{(1-\mu(A))(1-\mu(B))}-\sqrt{\mu(A)\mu(B)}
\end{eqnarray}
where we use the plus sign in front of $\sqrt{1-n^2}$ in the third coordinate of $|B\rangle$ in case $\mu(A)+\mu(B)\le1$, and the minus sign in case $1<\mu(A)+\mu(B)$.

\bigskip
\noindent
Let us work out these `context-interference' representations for the items and pairs of concepts of Tables 1 and 2 that could not be modeled by an `interference alone' representation, i.e. the items missing in Table 3. 

For example, consider the item {\it Refrigerator} with respect to the pair of concepts {\it House Furnishings} and {\it Furniture} and their disjunction {\it House Furnishings or Furniture}. We have $\mu(A)=0.9$, $\mu(B)=0.7$ and $\mu(A\ {\rm or}\ B)=0.575$. Hence we have $\mu(A\ {\rm or}\ B)<\mu(A)$ and $\mu(A\ {\rm or}\ B)<\mu(B)$, which means that we are in the situation of {\it double underextension}. We have $R=-0.6205$, $D_u=1.0103$, $n=0.7331$, $n'=0.8312$, $\beta=119.3535^\circ$, and $|A\rangle=(0.6955, 0.2318, 0.6801)$, $|B\rangle=e^{i119.3535^\circ}(0.6955, -0.4553, -0.5559)$. Let us consider the item {\it Almond} with respect to the pair of concepts {\it Fruits} and {\it Vegetables} and their disjunction {\it Fruits or Vegetables}. We have $\mu(A)=0.2$, $\mu(B)=0.1$ and $\mu(A\ {\rm or}\ B)=0.425$. Hence we have $\mu(A)<\mu(A\ {\rm or}\ B)$ and $\mu(B)<\mu(A\ {\rm or}\ B)$, which means that we are in a situation of {\it double overextension}. Additionally this is a type of overextension that cannot be modeled within a classical Kolmogorovian probability model, as we have analyzed in Aerts (2009), because $\mu(A)+\mu(B)=0.3<0.425=\mu(A\ {\rm or}\ B)$. For the $\compl^3$ model that we propose in this section we find $R=0.7071$, $D_o=1.45$, $n=0.7873$, $n'=0.7422$, $\beta=51.9269^\circ$, and $|A\rangle=(0.3521, 0.7042, 0.6166)$, $|B\rangle=e^{i51.9269^\circ}(0.2347, -0.7042, 0.6701)$. Consider next the item {\it Lineman's Flag} with respect to the pair of concepts {\it Sportswear} and {\it Sports Equipment}. We have $\mu(A)=0.1$, $\mu(B)=1$ and $\mu(A\ {\rm or}\ B)=0.75$. Hence we have $\mu(A)\le\mu(A\ {\rm or}\ B)\le\mu(B)$, which means that we are in a situation of convexity. For the $\real^2$ model that we propose in this section we find we find $R=-0.3162$, $D_c=0.225$, $n=0.6032$, $n'=0.9726$, $\beta=90^\circ$ and $|A\rangle=(0.1907, 0.5722, 0.7976)$ and $|B\rangle=e^{i90^\circ}(0.9726, 0, -0.2326)$.  

In Table 4 we have worked out the representations for all the items of Tables 1 and 2 that could not be modeled by an `interference alone' modeling, i.e. those that are absent from Table 3.

\begin{table}
\begin{center}
\scriptsize
{\begin{tabular}{lll}
\hline
 & $|A\rangle$ & $|B\rangle$ \\
\hline 
\multicolumn{3}{l}{\it $A$=House Furnishings, $B$=Furniture} \\
\hline
{\it Ashtray} & (0.5477, 0.3586, 0.7559) & $e^{i113.8762^\circ}$(0.5477, -0.8367, 0) \\
{\it Refrigerator} & (0.6955, 0.2318, 0.6801) & $e^{i119.3535^\circ}$(0.6955, -0.4553, -0.5559) \\
{\it Park Bench} & (0, 0.9334, 0.3588) & $e^{i90^\circ}$(0.2286, -0.3493, 0.9087) \\
{\it Waste-Paper Basket} & (0.4682, 0, 0.8836) & $e^{i90^\circ}$(0.6622, -0.6622, -0.35089) \\
{\it Sink Unit} & (0.6792, 0.2264, 0.6982) & $e^{i107.2101^\circ}$(0.6792, -0.5545, -0.4809) \\
\hline
\multicolumn{3}{l}{\it $A$=Hobbies, $B$=Games} \\
\hline
{\it Judo} & (0.5993, 0, 0.8005) & $e^{i90^\circ}$(0.7091, -0.4642, -0.5308) \\
{\it Discus Throwing} & (0.6772, 0, 0.7358) & $e^{i127.6699^\circ}$(0.6772, -0.3910, -0.6233) \\
{\it Karate} & (0.5993, 0, 0.8005) & $e^{i90^\circ}$(0.7091, -0.4642, -0.5308) \\
{\it Beer Drinking} & (0.8944, 0.4472, 0) & $e^{i90^\circ}$(0.3464, -0.6928, -0.6325)  \\
{\it Wrestling} & (0.2821, 0.0940, 0.9547) & $e^{i90^\circ}$(0.7641, -0.6239, -0.1643)   \\
\hline
\multicolumn{3}{l}{\it $A$=Spices, $B$=Herbs} \\
\hline
{\it Monosodium Glutamate} & (0.2967, 0.7063, 0.6427) & $e^{i47.3771^\circ}$(0.2354, -0.7063, 0.6676) \\
{\it Sugar} & (0, 0.7071, 0.7071) & $e^{i0^\circ}$(0, -0.7071, 0.7071) \\
{\it Sesame Seeds} & (0.5178, 0.7056, 0.4838) & $e^{i61.9894^\circ}$(0.5761, -0.7056, 0.4126)  \\
{\it Horseradish} & (0.3462, 0.6924, 0.6330) & $e^{i46.8499^\circ}$(0.5654, -0.6924, 0.4482)  \\
{\it Vanilla} & (0.5799, 0.4735, 0.6630) & $e^{i90^\circ}$(0, -0.8138, 0.5811)  \\
\hline
\multicolumn{3}{l}{\it $A$=Instruments, $B$=Tool} \\
\hline
{\it Pencil Eraser} & (0.6316, 0.7735, 0.0523) & $e^{i90^\circ}$(0.3737, -0.2446, -0.8947) \\
{\it Bicycle Pump} & (0.6972, 0, 0.7169) & $e^{i150.5017^\circ}$(0.6972, -0.2324, -0.6781) \\
{\it Computer} & (0.6924, 0.5654, 0.4482) & $e^{i101.1754^\circ}$(0.6924, -0.3462, -0.6330)\\
{\it Spoon} & (0.7693, 0.5645, 0.2993) & $e^{i90^\circ}$(0.4526, -0.1509, -0.8789) \\
\hline
\multicolumn{3}{l}{\it $A$=Pets, $B$=Farmyard Animals} \\
\hline
{\it Camel} & (0.3300, 0.4042, 0.8531) & $e^{i90^\circ}$(0, -0.9037, 0.4281) \\
{\it Guide Dog for the Blind} & (0.7938, 0.5196, 0.3161) & $e^{i29.4983^\circ}$(0, -0.5196, 0.8544) \\
{\it Spider} & (0.6866, 0.6866, 0.2393) & $e^{i76.6094^\circ}$(0.5038, -0.6866, 0.5242) \\
{\it Monkey} & (0.5412, 0.5412, 0.6436) & $e^{i90^\circ}$(0, -0.7654, 0.6436) \\
{\it Rat} &  (0.6800, 0.6800, 0.2741) & $e^{i111.3866^\circ}$(0.6800, -0.4452, -0.5826) \\
{\it Field Mouse} & (0.2823, 0.8468, 0.4508) & $e^{i90^\circ}$(0.7468, -0.4889, 0.4508) \\
\hline 
\multicolumn{3}{l}{\it $A$=Sportswear, $B$=Sports Equipment} \\
\hline
{\it Sunglasses} &  (0.4202, 0.5146, 0.7474) & $e^{i123.0845^\circ}$(0.4202, -0.8404, 0.3424) \\
{\it Lineman's Flag} & (0.1907, 0.5722, 0.7976) & $e^{i90^\circ}$(0.9726, 0, -0.2326) \\
{\it Bathing Costume} & (0.6847, 0, 0.7288) & $e^{i120^\circ}$(0.6847, -0.3424, -0.6434) \\
\hline
\multicolumn{3}{l}{\it $A$=Fruits, $B$=Vegetables} \\
\hline
{\it Parsley} & (0, 0.6847, 0.7288) & $e^{i32.0892^\circ}$(0.3424, -0.6847, 0.6434) \\
{\it Olive} & (0.6514, 0.6514, 0.3891) & $e^{i37.3569^\circ}$(0.2171, -0.6514, 0.7270) \\
{\it Broccoli} & (0, 0.4370, 0.8995) & $e^{i0^\circ}$(0.8740, -0.4370, 0.2123) \\
{\it Root Ginger} & (0, 0.6686, 0.7436) & $e^{i34.4444^\circ}$(0.4377, -0.6686, 0.6012) \\
{\it Tomato} & (0.7071, 0.4629, 0.5345) & $e^{i0^\circ}$(0.7071, -0.4629, -0.5345) \\
{\it Coconut} & (0.7938, 0.5196, 0.3161) & $e^{i0^\circ}$(0, -0.5196, 0.8544) \\
{\it Mushroom} & (0, 0.6180, 0.7862) & $e^{i19.1881^\circ}$(0.6180, -0.6180, 0.4859) \\
{\it Green Pepper} & (0.4111, 0.6280, 0.6607) & $e^{i50.7870^\circ}$(0.7692, -0.6280, 0.1183) \\
{\it Watercress} & (0, 0.5774, 0.8165) & $e^{i35.6591^\circ}$(0.7071, -0.5774, 0.4082) \\
{\it Garlic} & (0.2347, 0.7042, 0.6701) & $e^{i45.3817^\circ}$(0.3521, -0.7042, 0.6166) \\
{\it Yam} & (0.5305, 0.5865, 0.6120) & $e^{i51.1376^\circ}$(0.7993, -0.5865, -0.1308) \\
{\it Elderberry} & (1, 0, 0) & $e^{i90^\circ}$(0, -0.5, 0.8660) \\
{\it Almond} & (0.3521, 0.7042, 0.6166) & $e^{i51.9269^\circ}$(0.2347, -0.7042, 0.67012) \\
\hline
\multicolumn{3}{l}{\it $A$=Household Appliances, $B$=Kitchen Utensils} \\
\hline
{\it Cake Tin} &  (0.4448, 0.5448, 0.7109) & $e^{i30.1953^\circ}$(0.8322, -0.5448, 0.1032) \\
{\it Rubbish Bin} &  (0.7071, 0.7071, 0) & $e^{i53.1301^\circ}$(0.7071, -0.7071, 0) \\
{\it Electric Toothbrush} & (0.8400, 0.4200, 0.3436) & $e^{i90^\circ}$(0, -0.6331, 0.7740) \\
\hline
\end{tabular}}
\end{center}
\caption{`Contex and Interference' modeling of concepts and items of experiment 2 of \cite{hampton1988b}.}
\end{table}

\normalsize
The situation where $\mu(A)=\mu(B)=0$ and $0<\mu(A\ {\rm or}\ B)$ needs special attention. Indeed, in this situation the nominator of the right hand sight of (\ref{sol01}), i.e. expression $n^2\mu(A)+n'^2\mu(B)+2n'n'\sqrt{\mu(A)\mu(B)}\cos\phi$, equals 0, which means that, in case the denominator of the same expression is different from 0, equation (\ref{sol01}) forces $\mu(A\ {\rm or}\ B)$ to be 0. Moreover the situation does appear amongst \cite{hampton1988b}'s data. Indeed, for the item {\it Sugar} with respect to the pair of concepts {\it Spices} and {\it Herbs} and their disjunction {\it Spices or Herbs}, we have $\mu(A)=\mu(B)=0$ and $\mu(A\ {\rm or}\ B)=0.2$. We want to show in some detail how this situation gets solved in our model, since the solution shows us one of the powerful aspects of the approach.

First we remark that a solution exists for $0<\mu(A)\le\epsilon$ and $0<\mu(B)\le\epsilon$ for a number $\epsilon$ however small. The mathematical approach of such a situation consists in defining the situation for $\mu(A)=\mu(B)=0$ as the limiting case for $\epsilon\mapsto0$ of the situation encountered for $0<\mu(A)\le\epsilon$ and $0<\mu(B)\le\epsilon$. Let is prove that this works well for our model, and that we can indeed work out a representation for the item {\it Sugar} with respect to the pair of concepts {\it Spices} and {\it Herbs} and their disjunction {\it Spices or Herbs}. Let us take $\mu(A)=\mu(B)=\epsilon$ and calculate the different quantities involved. We have
\begin{equation}
R(\epsilon)=\sqrt{(1-\epsilon)(1-\epsilon)}-\sqrt{\epsilon\epsilon}=1-2\epsilon
\end{equation}
and
\begin{equation}
D_o(\epsilon)=(2-2\epsilon)^2-4(1-\epsilon)^2(1-(1-2\epsilon)^2)=4(1-\epsilon)^2(1-2\epsilon)^2 
\end{equation}
Further we have
\begin{eqnarray}
n(\epsilon)=n'(\epsilon)&=&\sqrt{{2(1-\epsilon)-2(1-\epsilon)(1-2\epsilon) \over 8\epsilon(1-\epsilon)^2}} = \sqrt{{2-2(1-2\epsilon) \over 8\epsilon(1-\epsilon)}} \nonumber \\
&=&\sqrt{{\epsilon \over 2\epsilon(1-\epsilon)}} = {1 \over \sqrt{2(1-\epsilon)}}
\end{eqnarray}
and
\begin{eqnarray}
\!\!\!\!\!\!\!\!\!\!\!\!\!\!\!\! \cos\phi(\epsilon)&=&{{1 \over 2(1-\epsilon)}(\mu(A\ {\rm or}\ B)-\epsilon)+{1 \over 2(1-\epsilon)}(\mu(A\ {\rm or}\ B)-\epsilon) \over 2{1 \over 2(1-\epsilon)}(1-\mu(A\ {\rm or}\ B))\epsilon+2{1 \over 2(1-\epsilon)}\mu(A\ {\rm or}\ B)\sqrt{1-2\epsilon+\epsilon^2}} \\
&=&{{\mu(A\ {\rm or}\ B)-\epsilon \over 1-\epsilon} \over {\epsilon \over 1-\epsilon}(1-\mu(A\ {\rm or}\ B))+{\sqrt{1-2\epsilon+\epsilon^2} \over 1-\epsilon}\mu(A\ {\rm or}\ B)} \\
&=&{\mu(A\ {\rm or}\ B)-\epsilon \over \epsilon(1-\mu(A\ {\rm or}\ B))+\sqrt{1-2\epsilon+\epsilon^2}\mu(A\ {\rm or}\ B)} \\
&=&{\mu(A\ {\rm or}\ B)-\epsilon \over \epsilon+(\sqrt{1-2\epsilon+\epsilon^2}-\epsilon)\mu(A\ {\rm or}\ B)}
\end{eqnarray}
and, substituting in (\ref{sol01}) we get
\begin{eqnarray}
&&\mu(A\ {\rm or}\ B)(\epsilon)= \\
&&{n^2(\epsilon)\epsilon+n'^2(\epsilon)\epsilon+2n(\epsilon)n'(\epsilon)\sqrt{\epsilon^2}\cos\phi(\epsilon) \over n^2(\epsilon)+n'^2(\epsilon)+2n(\epsilon)n'(\epsilon)\sqrt{\epsilon^2}\cos\phi(\epsilon)-2n(\epsilon)n'(\epsilon)\sqrt{(1-\epsilon)(1-\epsilon)}\cos\phi(\epsilon)} \nonumber \\ \label{righthandside}
&&={2n^2(\epsilon)\epsilon(1+\cos\phi(\epsilon)) \over 2n^2(\epsilon)+2n^2(\epsilon)\epsilon\cos\phi(\epsilon)-2n^2(\epsilon)\sqrt{1-2\epsilon+\epsilon^2}\cos\phi(\epsilon)}
\end{eqnarray}
This means that we have
\begin{eqnarray}
&&\lim_{\epsilon\mapsto0}n(\epsilon)=\lim_{\epsilon\mapsto0}n'(\epsilon)={1 \over \sqrt{2}} \\
&&\lim_{\epsilon\mapsto0}\cos\phi(\epsilon)=1 \\
&&\lim_{\epsilon\mapsto0}\mu(A\ {\rm or}\ B)(\epsilon)={0 \over 0}
\end{eqnarray}
which shows that indeed the limit for $\epsilon$ going to 0 of $\mu(A\ {\rm or}\ B)(\epsilon)$ is not well defined. We can however calculate it by using the necessary techniques from analysis to calculate the limit in a situation where substituting the value gives an undefinedness of the type ${0 \over 0}$. More specifically we use l'H\^opital's rule, which states we can take the derivative $d \over d\epsilon$ of nominator and denominator, and then calculate the limit of the fraction of these derivatives. Let us call $f(\epsilon)$ the right hand side of (\ref{righthandside}), and substitute $n(\epsilon)$ and $\cos\phi(\epsilon)$ in it. This gives
\begin{eqnarray}
f(\epsilon)&=&{2n^2(\epsilon)\epsilon(1+\cos\phi(\epsilon)) \over 2n^2(\epsilon)+2n^2(\epsilon)\epsilon\cos\phi(\epsilon)-2n^2(\epsilon)\sqrt{1-2\epsilon+\epsilon^2}\cos\phi(\epsilon)} \\
&=&{{\epsilon \over 1-\epsilon}(1+\cos\phi(\epsilon)) \over {1 \over 1-\epsilon}+{\epsilon \over 1-\epsilon}\cos\phi(\epsilon)-{1 \over 1-\epsilon}\sqrt{1-2\epsilon+\epsilon^2}\cos\phi(\epsilon)} \\
&=&{\epsilon(1+\cos\phi(\epsilon)) \over 1+\epsilon\cos\phi(\epsilon)-\sqrt{1-2\epsilon+\epsilon^2}\cos\phi(\epsilon)}
\end{eqnarray}
and hence
\begin{eqnarray}
&&\!\!\!\! \lim_{\epsilon\mapsto0}f(\epsilon)=\lim_{\epsilon\mapsto0}{\epsilon(1+\cos\phi(\epsilon)) \over 1+\epsilon\cos\phi(\epsilon)-\sqrt{1-2\epsilon+\epsilon^2}\cos\phi(\epsilon)} \\
&&\!\!\!\! =\lim_{\epsilon\mapsto0}{1+\cos\phi(\epsilon)+\epsilon{d \over d\epsilon}\cos\phi(\epsilon) \over \cos\phi(\epsilon)+\epsilon{d \over d\epsilon}\cos\phi(\epsilon)-{-1+\epsilon \over \sqrt{1-2\epsilon+\epsilon^2}}\cos\phi(\epsilon)-\sqrt{1-2\epsilon+\epsilon^2}{d \over d\epsilon}\cos\phi(\epsilon)} \nonumber \\
&&\!\!\!\! ={2 \over 2-\lim_{\epsilon\mapsto0}{d \over d\epsilon}\cos\phi(\epsilon)}
\end{eqnarray}
and after calculation we find
\begin{eqnarray}
\lim_{\epsilon\mapsto0}{d \over d\epsilon}\cos\phi(\epsilon)&=&{-\mu(A\ {\rm or}\ B)-\mu(A\ {\rm or}\ B)(1-2\mu(A\ {\rm or}\ B)) \over \mu(A\ {\rm or}\ B)^2} \\
&=&{-2+2\mu(A\ {\rm or}\ B) \over \mu(A\ {\rm or}\ B)}
\end{eqnarray}
This shows that 
\begin{eqnarray}
\lim_{\epsilon\mapsto0}f(\epsilon)&=&{2 \over 2-({-2+2\mu(A\ {\rm or}\ B) \over \mu(A\ {\rm or}\ B)})}={2 \over {2\mu(A\ {\rm or}\ B)-(-2+2\mu(A\ {\rm or}\ B)) \over \mu(A\ {\rm or}\ B)}} \\
&=&{2\mu(A\ {\rm or}\ B) \over 2\mu(A\ {\rm or}\ B)-(-2+2\mu(A\ {\rm or}\ B))}=\mu(A\ {\rm or}\ B)
\end{eqnarray}
which proves that our theory models the correct value of $\mu(A\ {\rm or}\ B)$.

Returning to the modeling of the item {\it Sugar} with respect to the pair of concepts {\it Spices} and {\it Herbs} and their disjunction {\it Spices or Herbs}, we get for the vectors $|A\rangle$ and $|B\rangle$ representing the concepts {\it Spices} and {\it Herbs} the following: $|A\rangle=(0, 0.7071, 0.7071)$ and $|B\rangle=e^{i0^\circ}(0, -0.7071, 0.7071)$.

\subsection{Graphical Representation of the Two-Concept Situation}
We can make a graphical representation of the three situations, double underextension, convexity and double overextension, by representing the normalized projections of the vectors $|A\rangle$, $|B\rangle$ and ${1 \over \sqrt{2}}(|A\rangle+|B\rangle)$ on the plane $N({\compl^3})$ of the three dimensional complex representation that we worked out in section \ref{compl3representation}. Hence, more specifically we represent the vectors $|A^N\rangle$, $|B^N\rangle$ and $|AB^N\rangle$ we introduced in (\ref{Nvectors}).

\begin{figure}[H]
\centerline {\includegraphics[width=8cm]{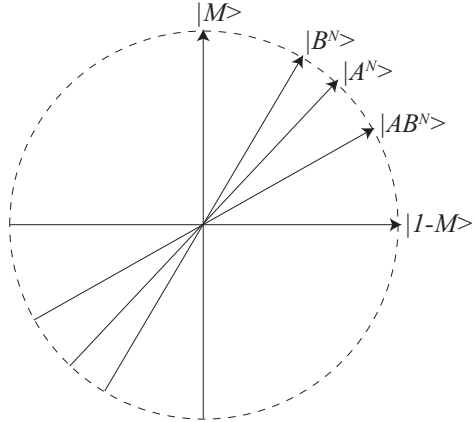}}
\caption{A graphical representation of a situation of underextension. We have $|\langle M|AB^N\rangle|^2\le|\langle M|A^N\rangle|^2$ and $|\langle M|AB^N\rangle|^2\le|\langle M|B^N\rangle|^2$.}
\label{fig1}
\end{figure}

Figure \ref{fig1} graphically represents a situation of underextension.

Hence, membership weight $\mu(A\ {\rm or}\ B)$, given by the square of the cosinus of the angle between vector $|AB^N\rangle$ and vector $|M\rangle$, is smaller than membership weight $\mu(A)$, given by the square of the cosinus of the angle between vector $|A^N\rangle$ and vector $|M\rangle$, and also smaller than membership weight $\mu(B)$, given by the square of the cosinus of the angle between vector $|B^N\rangle$ and vector $|M\rangle$.

\begin{figure}[H]
\centerline {\includegraphics[width=8cm]{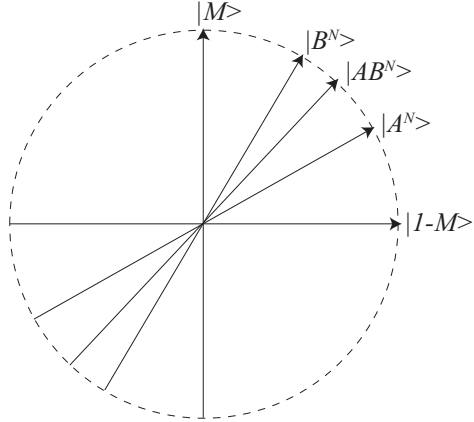}}
\caption{A graphical representation of a situation of convexity. We have $|\langle M|A^N\rangle|^2\le|\langle M|AB^N\rangle|^2$ and $|\langle M|AB^N\rangle|^2\le|\langle M|B^N\rangle|^2$.}
\label{fig2}
\end{figure}

Figure \ref{fig2} graphically represents a situation of convexity.

Hence, membership weight $\mu(A\ {\rm or}\ B)$, given by the square of the cosinus of the angle between the vector $|AB^N\rangle$ and the vector $|M\rangle$, is bigger than the membership weight $\mu(A)$, given by the square of the cosinus of the angle between the vector $|A^N\rangle$ and the vector $|M\rangle$, and smaller than the membership weight $\mu(B)$, given by the square of the cosinus of the angle between the vector $|B^N\rangle$ and the vector $|M\rangle$.

Figure 3 graphically represents a situation of overextension. Hence, membership weight $\mu(A\ {\rm or}\ B)$, given by the square of the cosinus of the angle between the vector $|AB^N\rangle$ and the vector $|M\rangle$, is bigger than the membership weight $\mu(A)$, given by the square of the cosinus of the angle between the vector $|A^N\rangle$ and the vector $|M\rangle$, and also bigger than the membership weight $\mu(B)$, given by the square of the cosinus of the angle between the vector $|B^N\rangle$ and the vector $|M\rangle$.

\begin{figure}[H]
\centerline {\includegraphics[width=8cm]{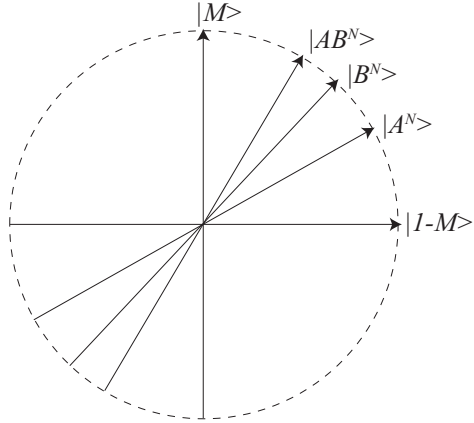}}
\caption{A graphical representation of a situation of overextension. We have $|\langle M|A^N\rangle|^2\le|\langle M|AB^N\rangle|^2$ and $|\langle M|B^N\rangle|^2\le|\langle M|AB^N\rangle|^2$.}
\label{fig3}
\end{figure}

\section{Conclusion}
We introduced a quantum model for the description of the disjunction of two concepts that is able to model the outcomes of experiments that measure the membership weight of items with respect to the disjunction of two concepts in function of the membership weight of these items with respect to each of the concepts apart. The phenomenon of quantum interference accounts for the deviations from a classical interpretation of the disjunction as measured in experiments in \cite{hampton1988b}, and is called underextension or overextension, depending on whether the disjunction weight is smaller or bigger than expected classically. A considerable number of the deviations from the classical disjunction -- underextension as well as overextension -- are bigger in size than the maximum that can be accounted for by quantum interference if the concepts are considered to be represented by orthogonal quantum states. If we additionally introduce the quantum phenomenon of `contextual influence', these deviations too can be modeled by the quantum mechanical formalism. Introducing the `influence of contexts' as an effect of the `act of choosing the specific item for which the membership weight is measured' produces quantum states for the considered concepts that are no longer orthogonal after the context effect has changed the states of the concepts. To further elaborate and determine the meaning of the parameters appearing in the `context and interference quantum model', additional experiments could be performed of the type considered in \cite{aertsgabora2005a}. The quantum model which introduces the influence of context as well as interference entails a simple graphical illustration of the effects of underextension and overextension. 
In this article we only treat the situation of the disjunction of concepts. The theory's full power will show when we use it also for the modeling of the situation of the conjunction of concepts and for the elaboration of a fundamental mechanism for the formation of concepts in future work.

The material of the present article was first elaborated in the summer of 2006, at the time as part of the ongoing activity of our Brussels research group of investigating the use of quantum structures to model entities and interactions in domains of science different from the micro-world \cite{aertsaerts1994,aertsbroekaertsmets1999a,aertsbroekaertsmets1999b,aertsaertsbroekaertgabora2000,gaboraaerts2002a,gaboraaerts2002b,aertsczachorgaborakunaposiewnikpykaczsyty2003,aertsczachor2004,aertsgabora2005a,aertsgabora2005b,aertsczachordhooghe2006}. Now, three years later, this period turns out to have been a starting phase for considerable parallel research activity by different scientists on this idea of applying quantum structures in other domains of science than the micro-world. Three workshops have since been organized on this research aim, which is now also commonly referred to as  `quantum interaction' \cite{QI2007,QI2008,QI2009}, while a growing international group of researchers is becoming very actively engaged in it.

So far, the application of quantum structures has been focused mainly on economics \cite{schaden2002,baaquie2004,haven2005,bagarello2006}, psychology and cognition, concepts theories and decision theory \cite{aertsaerts1994,gaboraaerts2002a,gaboraaerts2002b,aertsgabora2005a,aertsgabora2005b,busemeyerwangtownsend2006}, and language, information retrieval and artificial intelligence \cite{widdows2003,widdows2006,widdowspeters2003,aertsczachor2004,vanrijsbergen2004,bruzacole2005,aertsczachordhooghe2006}.


\begin{thebibliography}{99}
\setlength{\itemsep}{-1.5mm}
\bibitem{hampton1991} J. A. Hampton,  The combination of prototype concepts.  In P. Schwanenflugel (Ed.), {\it The Psychology of Word Meanings}. Hillsdale, NJ: Erlbaum, (1991).

\bibitem{hampton1997a} J. A. Hampton, Conceptual combination: Conjunction and negation of natural concepts. {\it Memory \& Cognition} {\bf 25}, 888-909 (1997).

\bibitem{hampton1997b} J. A. Hampton, Conceptual combination. In K. Lamberts \& D. Shanks (Eds.), {\it Knowledge, Concepts, and Categories}. Hove: Psychology Press, 133-159 (1997).

\bibitem{kundamillerclaire1990} Z. Kunda, D. T. Miller \& T. Claire, Combining social concepts: The role of causal reasoning. {\it Cognitive Science} {\bf 14}, 551-577 (1990).

\bibitem{oshersonsmith1981} D. N. Osherson \& E. E. Smith, On the adequacy of prototype theory as a theory of concepts. {\it Cognition} {\bf 9}, 35-58  (1981). 

\bibitem{oshersonsmith1982} D. N. Osherson \& E. E. Smith, Gradedness and conceptual combination. {\it Cognition} {\bf 12}, 299-318 (1982).

\bibitem{rips1995}  L. J. Rips, The current status of research on concept combination. {\it Mind \& Language} {\bf 10}, 72-104 (1995).

\bibitem{smithosherson1984} E. E. Smith \& D. N. Osherson, Conceptual combination with prototype concepts. {\it Cognitive Science} {\bf 8}, 357-361  (1984). 

\bibitem{smithoshersonripskeane1988} E. E. Smith,  D. N. Osherson, L. J. Rips \& M. Keane, Combining prototypes: A selective modification model.  {\it Cognitive Science} {\bf 12}, 485-527 (1988).

\bibitem{springermurphy1992} K. Springer \& G. L. Murphy, Feature availability in conceptual combination. {\it Psychological Science} {\bf 3}, 111-117 (1992). 

\bibitem{hampton1988a} J. A. Hampton, Overextension of conjunctive concepts: Evidence for a unitary model for concept typicality and class inclusion. {\it Journal of Experimental Psychology: Learning, Memory, and Cognition} {\bf 14}, 12-32 (1988).

\bibitem{hampton1988b} J. A. Hampton, Disjunction of natural concepts. {\it Memory \& Cognition} {\bf 16}, 579-591 (1988).

\bibitem{hampton1996} J. A. Hampton, Conjunctions of visually-based categories: overextension and compensation. {\it Journal of Experimental Psychology: Learning, Memory and Cognition} {\bf 22}, 378-396 (1996).

\bibitem{stormsdeboeckvanmechelengeeraerts1993}  G. Storms, P. De Boeck, I. Van Mechelen \& D. Geeraerts, Dominance and non-commutativity effects in concept conjunctions: Extensional or intensional basis? {\it Memory \& Cognition} {\bf 21}, 752-762  (1993).

\bibitem{stormsdeboeckvanmechelenruts1996} G. Storms, P. De Boeck, I. Van Mechelen \&  W. Ruts, The dominance effect in concept conjunctions: Generality and interaction aspects. {\it Journal of Experimental Psychology: Learning, Memory \& Cognition} {\bf 22}, 1-15 (1996).

\bibitem{stormsrutsvandenbroucke1998} G. Storms, W. Ruts \& A. Vandenbroucke, Dominance, overextensions, and the conjunction effect in different syntactic phrasings of concept conjunctions. {\it European Journal of Cognitive Psychology} {\bf 10}, 337-372 (1998).

\bibitem{rosch1973a}  E. Rosch, Natural categories. {\it Cognitive Psychology} {\bf 4}, 328-350 (1973).

\bibitem{rosch1973b}  E. Rosch, On the internal structure of perceptual and semantic categories. In T. E. Moore (Ed.), {\it Cognitive Development and the Acquisition of Language}. New York: Academic Press, (1973). 

\bibitem{rosch1975}  E. Rosch, Cognitive representations of semantic categories. {\it Journal of Experimental Psychology: General} {\bf 104}, 192-232 (1975). 

\bibitem{rosch1978} E. Rosch, Principles of categorization. In E. Rosch \& B. Lloyd (Eds.), {\it Cognition and Categorization}. Hillsdale, NJ: Lawrence Erlbaum, 133-159 (1978). 

\bibitem{rosch1983} E. Rosch, Prototype classification and logical classification: The two systems. In E. K. Scholnick (Ed.), {\it New Trends in Conceptual Representation: Challenges to PiagetÕs Theory?}. Hillsdale, NJ: Lawrence Erlbaum, 133-159 (1983).

\bibitem{zadeh1965} L. Zadeh, Fuzzy sets. {\it Information \& Control} {\bf 8}, 338-353 (1965).

\bibitem{dirac1958} P. A. M. Dirac, {\it Quantum mechanics}, 4th ed. London: Oxford University Press, (1958).

\bibitem{aertsgabora2005a} D. Aerts \&  L. Gabora, A theory of concepts and their combinations I: The structure of the sets of contexts and properties. {\it Kybernetes} {\bf 34}, 167-191 (2005).

\bibitem{aertsgabora2005b} D. Aerts \&  L. Gabora, A theory of concepts and their combinations II: A Hilbert space representation. {\it Kybernetes} {\bf 34}, 192-221 (2005).

\bibitem{aertsaerts1994} D. Aerts \&  S. Aerts, Applications of quantum statistics in psychological studies of decision processes. {\it Foundations of Science} {\bf 1}, 85-97 (1994). Reprinted in B. Van Fraassen (Eds.), {\it Topics in the Foundation of Statistics}. Dordrecht: Kluwer Academic, 111-122 (1997).

\bibitem{aertsbroekaertsmets1999a}  D. Aerts, J. Broekaert \& S. Smets, The liar paradox in a quantum mechanical perspective. {\it Foundations of Science} {\bf 4}, 115-132 (1999).

\bibitem{aertsbroekaertsmets1999b}  D. Aerts,  J. Broekaert \&  S. Smets, A quantum structure description of the liar paradox. {\it International Journal of Theoretical Physics} {\bf 38}, 3231-3239 (1999). 

\bibitem{aertsaertsbroekaertgabora2000}  D. Aerts,  S. Aerts, J. Broekaert \&  L. Gabora, The violation of Bell inequalities in the macroworld. {\it Foundations of Physics} {\bf 30}, 1387-1414 (2000). 

\bibitem{gaboraaerts2002a} L. Gabora \&  D. Aerts, Contextualizing concepts. In {\it Proceedings of the 15th International FLAIRS Conference. Special track: Categorization and Concept Representation: Models and Implications}, Pensacola Florida, May 14-17, American Association for Artificial Intelligence, 148-152 (2002). 

\bibitem{gaboraaerts2002b} L. Gabora \&  D. Aerts, Contextualizing concepts using a mathematical generalization of the quantum formalism. {\it Journal of Experimental and Theoretical Artificial Intelligence} {\bf 14}, 327-358 (2002). Preprint at http://arXiv.org/abs/quant-ph/0205161 

\bibitem{aertsczachorgaborakunaposiewnikpykaczsyty2003}  D. Aerts, M. Czachor, L. Gabora, M. Kuna, A. Posiewnik,   J. Pykacz \&  M. Syty, Quantum morphogenesis: A variation on Thom's catastrophe theory. {\it Physical Review E} {\bf 67}, 051926 (2003). 

\bibitem{aertsczachor2004} D. Aerts \&  M. Czachor, Quantum aspects of semantic analysis and symbolic artificial intelligence. {\it Journal of Physics A, Mathematical and Theoretical} {\bf 37}, L123-L132 (2004).

\bibitem{aertsczachordhooghe2006}  D. Aerts, M. Czachor \&  B. D'Hooghe, Towards a quantum evolutionary scheme: violating BellÕs inequalities in language. In N. Gontier, J. P. Van Bendegem, \& D. Aerts (Eds.), {\it Evolutionary Epistemology, Language and Culture - A Non Adaptationist Systems Theoretical Approach}. Dordrecht: Springer, (2006).

\bibitem{QI2007} {\it AAAI Spring Symposium: Quantum Interaction}. Stanford University, California, March 26-28 (2007). Web Reference http://ir.dcs.gla.ac.uk/qi2007/

\bibitem{QI2008} {\it Second International Symposium QI 2008: Quantum Interaction}. Oxford University, United Kingdom, March 26-28 (2008). Web Reference: http://ir.dcs.gla.ac.uk/qi2008/

\bibitem{QI2009} {\it Third International Symposium QI 2009: Quantum Interaction}. German Research Center for Artificial Intelligence (DFKI), Saarbruecken, Germany, March 25-27 (2009). Web Reference: http://www-ags.dfki.uni-sb.de/~klusch/qi2009/

\bibitem{schaden2002} Schaden, M. (2002). Quantum finance: A quantum approach to stock price fluctuations. {\it Physica A, 316}, 511. 

\bibitem{baaquie2004} Baaquie, B. E. (2004). {\it Quantum Finance: Path Integrals and Hamiltonians for Options and Interest Rates}. Cambridge UK: Cambridge University Press.

\bibitem{haven2005} Haven, E. (2005). Pilot-wave theory and financial option pricing. {\it International Journal of Theoretical Physics, 44}, 1957-1962.

\bibitem{bagarello2006} Bagarello, F. (2006). An operatorial approach to stock markets. {\it Journal of Physics A, 39}, 6823-6840. 

\bibitem{busemeyerwangtownsend2006} Busemeyer, J. R., Wang, Z., \& Townsend, J. T. (2006). Quantum dynamics of human decision making. {\it Journal of Mathematical Psychology, 50}, 220-241.


\bibitem{widdows2003} Widdows, D. (2003). Orthogonal negation in vector spaces for modelling word-meanings and document retrieval. In {\it Proceedings of the 41st Annual Meeting of the Association for Computational Linguistics} (pp. 136-143). Sapporo, Japan, July 7-12. 

\bibitem{widdows2006} Widdows, D. (2006). {\it Geometry and Meaning}. CSLI Publications: University of Chicago Press. 

\bibitem{widdowspeters2003} Widdows, D., \& Peters, S. (2003). Word vectors and quantum logic: Experiments with negation and disjunction. In {\it Mathematics of Language 8} (pp. 141-154). Indiana: Bloomington.


\bibitem{vanrijsbergen2004} Van Rijsbergen, K. (2004). {\it The Geometry of Information Retrieval}. Cambridge UK: Cambridge University Press. 

\bibitem{bruzacole2005} Bruza, P. D. and Cole, R. J. (2005). Quantum logic of semantic space: An exploratory investigation of context effects in practical reasoning. In S. Artemov, H. Barringer, A. S. d'Avila Garcez, L.C. Lamb, J. Woods (Eds.) {\it We Will Show Them: Essays in Honour of Dov Gabbay}. College Publications.

\end{thebibliography}
\end{document}